\documentclass[a4paper,11pt]{article}
\pdfoutput=1 
\usepackage{jheppub} 
\usepackage[T1]{fontenc} 
\usepackage{float}
\usepackage{multirow}
\usepackage{subfig}
\usepackage{amsmath}
\usepackage[dvipsnames]{xcolor}
\usepackage{hyperref}
\usepackage[normalem]{ulem}
\usepackage{xspace}
\usepackage{comment}

\newcommand{\gambit}[0]{\textsf{GAMBIT}\xspace}

\title{
Global fit to the 2HDM with generic sources of flavour violation using \textsf{GAMBIT}
}
\author[a]{Peter Athron,}
\author[b,c]{Andreas Crivellin,}
\author[d]{Tom\'as E. Gonzalo,}
\author[e,f,g]{Syuhei Iguro}
\author[a]{and Cristian Sierra}

\affiliation[a]{Department of Physics and Institute of Theoretical Physics, Nanjing Normal University, Wenyuan Road, Nanjing, Jiangsu, 210023, China}
\affiliation[b]{Physik-Institut, Universität Zürich, Winterthurerstrasse 190, CH–8057 Zürich, Switzerland}
\affiliation[c]{PSI Center for Neutron and Muon Sciences, 5232 Villigen PSI, Switzerland}
\affiliation[d]{Institut für Theoretische Teilchenphysik, Karlsruher Institut für Technologie (KIT), D-76128 Karlsruhe, Germany}
\affiliation[e]{Institute for Advanced Research (IAR), Nagoya University, Nagoya 464--8601, Japan}
\affiliation[f]{Kobayashi-Maskawa Institute (KMI) for the Origin of Particles and the Universe,\\ Nagoya University, Nagoya 464--8602, Japan}
\affiliation[g]{KEK Theory Center, IPNS, KEK, Tsukuba 305--0801, Japan}

\emailAdd{peter.athron@coepp.org.au,andreas.crivellin@psi.ch,tomas.gonzalo@kit.edu,
igurosyuhei@gmail.com, cristian.sierra@njnu.edu.cn}

\keywords{Flavour physics phenomenology, two-Higgs doublet model, charged and neutral flavour anomalies, rare decays, global fit}

\preprint{PSI-PR-24-20, ZU-TH 50/24, \begin{flushright}TTP24-035, KEK--TH--2657
\end{flushright}}

\abstract{
We perform a global statistical analysis of the two-Higgs-doublet model with generic sources of flavour violation using \textsf{GAMBIT}. This is particularly interesting in light of deviations from the Standard Model predictions observed in $b\to c\tau\bar \nu$ and $b\to s\ell^+\ell^-$ transitions as well as the indications for a charged Higgs with a mass of 130\,GeV in top quark decays. Including all relevant constraints from precision, flavour and collider observables, we find that it is possible to simultaneously explain both the charged and neutral current $B$ anomalies.
We study the impact of using different values for the $W$ mass and the Standard Model prediction for $g-2$ of the muon and provide predictions for observables that can probe our model in the future such as lepton flavour violation searches at Belle II and Higgs coupling strength measurements at the high-luminosity LHC.}

\begin{document} 
\maketitle
\flushbottom

\newpage
\section{Introduction} \label{sec:introduction}
The Standard Model (SM) is extremely successful in describing the interactions of matter at sub-atomic scales~\cite{ParticleDataGroup:2024cfk}. However, several statistically significant deviations from the SM predictions, called anomalies, exist~\cite{Crivellin:2023zui}. In particular, the long-standing anomalies in semi-leptonic $B$ meson decays~\cite{Capdevila:2023yhq}, both in $b\to c \tau\overline{\nu}$~\cite{HFLAV:2022esi} transitions, i.e.~$R(D)$ and $R(D^*)$, ($3.3\sigma$) and in $b\to s\ell^+\ell^-$~observables ($\approx6\sigma$) persist; see Ref.~\cite{Alguero:2023jeh} for an overview. These observables point towards new physics (NP) and motivate the study of NP models capable of providing a combined explanation. Furthermore, there is a 3$\,\sigma$ excess in the exotic top decay $t\to b (H^+\to \bar{b}c)$~\cite{ATLAS:2023bzb} which motivates an extension of the scalar sector.

In this article, we investigate the possibility of a NP explanation of these anomalies within the context of the two-Higgs-doublet model (2HDM)~\cite{ Gunion:2002zf,Branco:2011iw} -- one of the simplest and most studied extensions of the SM scalar sector. The most general version with generic Yukawa couplings (G2HDM)\footnote{Sometimes this is also referred to as the type III 2HDM in the literature.} can explain $b\to c\tau\nu$ data at the $1\sigma$ level~\cite{Crivellin:2012ye,Crivellin:2013wna,Cline:2015lqp,Crivellin:2015hha,Lee:2017kbi,Iguro:2017ysu,Martinez:2018ynq,Fraser:2018aqj,Iguro:2018fni,Athron:2021auq,Iguro:2022uzz,Blanke:2022pjy,Ezzat:2022gpk,Fedele:2022iib, Das:2023gfz,Crivellin:2023sig} and address the anomalies in $b\to s\ell^+\ell^-$ transitions~\cite{Iguro:2018qzf, Crivellin:2019dun,Kumar:2022rcf,Iguro:2023jju, Crivellin:2023sig}, even though reaching the preferred central value of the latter is difficult. To evade constraints from collider searches, $B_s-\bar B_s$ mixing etc, only quite small regions of the parameter space remain valid and only particular benchmark points have been examined, while a combined statistical analysis of all available data and an identification of the allowed parameter space is still missing.\footnote{Efforts in this direction have been presented in the quark~\cite{Herrero-Garcia:2019mcy} and lepton sectors~\cite{Athron:2021auq}, separately.} Furthermore, contributions to other precision observables such as $g-2$ of the muon ($(g-2)_\mu$) and the $W$ mass ($m_W$) are in general expected in 2HDMs and have to be included in a global statistical analysis, even though the experimental and theoretical situation is not conclusive in these cases, as we will discuss in detail later. 

Such a global statistical analysis is the aim of this article. For this, we extend the work of Refs.~\cite{Athron:2021auq,Crivellin:2023sig} by including for instance recent measurements of the charged lepton flavour violating (cLFV) search in $t\to\mu\tau q$ decays from ATLAS~\cite{ATLAS:2024njy} and the latest universality test update from Belle II on $|g_{\mu}/g_e|$~\cite{Corona:2024nnn}. In addition to extending the set of observables and updating the data, we allow for additional Yukawa couplings to be non-zero which were previously not studied. For instance, an additional charm quark Yukawa coupling could enhance the effect in $b\to s\ell^+\ell^-$~\cite{Crivellin:2023sig}. We perform this global fit using the inference package \gambit, the Global And Modular Beyond-the-Standard-Model Inference Tool~\cite{Athron:2017ard, grev}, which is an open-source code in {\tt C++} to calculate observables and likelihoods for generic beyond the Standard Model (BSM) theories utilising different modules and external packages (see section~\ref{sec:Results} for more details).

The paper is organised as follows: In section~\ref{sec:GTHDM} we introduce the G2HDM model and in section~\ref{sec:Observables} we list the relevant observables, including flavour, collider and precision observables and afterwards present the results of the global fit and predictions for future experiments in section~\ref{sec:Results}. Finally, we conclude in section~\ref{sec:Conclusions} and show the potential impact of a sizable NP effect in $g-2$ of the muon on the fit in the appendix.

\section{The General Two Higgs Doublet Model} \label{sec:GTHDM}

The most general renormalisable scalar potential respecting gauge invariance is~\cite{Branco:2011iw,Gunion:2002zf}
\begin{alignat}{1}
V(\Phi_{1},\Phi_{2})=\: & m_{11}^{2}(\Phi_{1}^{\dag}\Phi_{1})+m_{22}^{2}(\Phi_{2}^{\dag}\Phi_{2})-m_{12}^{2}(\Phi_{1}^{\dag}\Phi_{2}+\Phi_{2}^{\dag}\Phi_{1})\nonumber \\
 & +\frac{1}{2}\lambda_{1}(\Phi_{1}^{\dag}\Phi_{1})^{2}+\frac{1}{2}\lambda_{2}(\Phi_{2}^{\dag}\Phi_{2})^{2}+\lambda_{3}(\Phi_{1}^{\dag}\Phi_{1})(\Phi_{2}^{\dag}\,\Phi_{2})+\lambda_{4}(\Phi_{1}^{\dag}\Phi_{2})(\Phi_{2}^{\dag}\Phi_{1})\nonumber \\
 & +\left(\frac{1}{2}\lambda_{5}(\Phi_{1}^{\dag}\Phi_{2})^{2}+\left(\lambda_{6}(\Phi_{1}^{\dag}\Phi_{1})+\lambda_{7}(\Phi_{2}^{\dag}\Phi_{2})\right)(\Phi_{1}^{\dag}\Phi_{2})+{\rm ~h.c.}\right) ,
 \label{eq:HiggsPotential}
\end{alignat}
where the parameters $m_{11}^2,\,m_{22}^2$ and $\lambda_{1-4}$ are real numbers (from hermiticity), whereas the  $\lambda_{5,6,7}$ and $m_{12}^2$ can in general be complex. For a CP-conserving potential, {which we assume in the following,} all parameters in Eq.~(\ref{eq:HiggsPotential}) are real and the total number of free parameters will be reduced from 14 to 10. Note that in our discussion of the flavour observables, only the resulting mixing angles among the scalars and their masses are relevant. We will come back to this point at the beginning of section~\ref{sec:modelparameters}.

Once the two scalars develop non-zero vacuum expectation values (VEVs) $\upsilon_1$ and $\upsilon_2$, the electroweak symmetry of the SM is spontaneously broken and the doublets are decomposed into components as
\begin{equation}
\ensuremath{\Phi_{i}=\left(\begin{array}{c}
\phi_{i}^{+}\\
\frac{1}{\sqrt{2}}(\upsilon_{i}+\rho_{i}+i\eta_{i})
\end{array}\right)},\quad i=1,2.\label{eq:Higgs_doublets}
\end{equation}
Linear combinations of the fields $\rho_i$, $\eta_i$ and $\phi_{i}^{\pm}$ form mass eigenstates
\begin{equation}
\begin{pmatrix}G_{Z}\\A
\end{pmatrix}=R_{\beta}\begin{pmatrix}\eta_{1}\\
\eta_{2}
\end{pmatrix},\quad\begin{pmatrix}G_{W^{\pm}}\\
H^{\pm}
\end{pmatrix}=R_{\beta}\begin{pmatrix}\phi_{1}^{\pm}\\
\phi_{2}^{\pm}
\end{pmatrix},\quad\begin{pmatrix}H\\
h
\end{pmatrix}=R_{\alpha}\begin{pmatrix}\rho_{1}\\
\rho_{2}
\end{pmatrix},\label{eq:rotation_matrices}
\end{equation}
where $\phi_{i}^{+}$ are electrically charged complex scalars and $\eta_i$ and $\rho_i$ neutral real scalars. $G_{W^{\pm}}$ and $G_{Z}$ correspond to longitudinal components of the $W$ and $Z$ bosons, while $h$ (the SM-like Higgs) and $H$ are physical CP-even states, $A$ a CP-odd state and $H^{\pm}$
is a charged Higgs boson. The rotation matrices are defined
as
\begin{equation} \label{eqn:GenericMixingMatrix}
R_{\theta}=\left(\begin{array}{cc}
\cos\theta & \sin\theta\\
-\sin\theta & \cos\theta
\end{array}\right),
\end{equation}
where $\theta$ is either $\alpha$ or $\beta$. The angle $\alpha$ is the mixing angle of the CP-even states, whereas the rotation angle $\beta$ is determined by
\begin{equation}
\tan\beta\equiv t_{\beta}=\frac{s_{\beta}}{c_{\beta}}=\frac{\upsilon_{2}}{\upsilon_{1}},
\end{equation}
with $\{s{_\beta}, c_\beta\} =\{\sin\beta, \cos\beta\}$ and $\upsilon_{2}^{2}+\upsilon_{1}^{2}=\upsilon^{2}$ with $\upsilon=246$~GeV being the SM VEV. The angle defined by $\beta-\alpha$ is the mixing angle between the CP-even Higgs mass eigenstates relative to the Higgs basis and the limit $\sin(\beta-\alpha)\equiv s_{\beta\alpha}\to 1$ is known as the alignment limit in which $h$ has the same properties as the SM Higgs. 

The most general Yukawa Lagrangian in the basis $\{\Phi_{1},\Phi_{2}\}$
reads~\cite{Branco:2011iw}
\begin{equation}
    -\mathcal{L}_{Yukawa}=\bar{Q}^{0}\,(Y^{1}_{u}\tilde{\Phi}_{1}+Y^{2}_{u}\tilde{\Phi}_{2})u_{{\rm R}}^{0}+\bar{Q}^{0}\,(Y^{1}_{d}\Phi_{1}+Y^{2}_{d}\Phi_{2})d_{{\rm R}}^{0}+\bar{L}^{0}\,(Y^{1}_{l}\Phi_{1}+Y^{2}_{l}\Phi_{2})l_{{\rm R}}^{0}+{\rm ~h.c.}\,\label{eq:yuk2d},
\end{equation}
where the superscript ``0'' notation refers to the flavour eigenstates, and $\tilde{\Phi}_j = i \sigma_2 \Phi_j^*$. The fermion mass matrices are determined by
\begin{equation}
    M_{f}=\frac{1}{\sqrt{2}}(v_{1}Y^{1}_{f}+v_{2}Y^{2}_{f}),\qquad f=u,d,l\, ,\label{masa-fermiones} 
\end{equation}
and are in general non-diagonal. Via a bi-unitary transformation
\begin{equation}
\bar{M}_{f}=V_{fL}^{\dagger}M_{f}V_{fR},\label{masa-diagonal}
\end{equation}
the mass eigenstates for the fermions are given by
\begin{equation}
    u=V_{u}^{\dagger}u^{0},\qquad d=V_{d}^{\dagger}d^{0},\qquad l=V_{l}^{\dagger}l^{0},\label{redfields}
\end{equation}
with
\begin{equation}
    \bar{M}_{f}=\frac{1}{\sqrt{2}}(v_{1}\tilde{Y}^{1}_{f}+v_{2}\tilde{Y}^{2}_{f}),\label{diag-Mf}
\end{equation}
where $\tilde{Y}^{i}_{f}=V_{fL}^{\dagger}Y^{i}_{f}V_{fR}$. We shall drop the tilde from now on. Solving for $Y^{1}_{f}$ we have
\begin{equation}
Y^{1,ba}_{f}=\frac{\sqrt{2}}{v\cos\beta}\bar{M}_{f}^{ba}-\tan\beta Y^{2,ba}_{f},
\end{equation}
and can write the Yukawa Lagrangian in the mass basis as
\begin{equation}
\begin{aligned}-\mathcal{L}_{Yukawa}\, =& \bar{u}_{b} \left(V_{bc}\rho^{ca}_{d}P_{R} - V_{ca}\rho^{cb*}_{u}P_{L}\right) d_{a}\,H^{+} + \bar{\nu}_{b}\rho^{ba}_{\ell}P_{R}l_{a}\,H^{+} + \mathrm{h.c.}\\
 & +\sum_{f=u,d,\ell}\sum_{\phi=h,H,A}\bar{f}_{b} \Gamma^{\phi ba}_{f}P_{R}f_{a}\phi+\mathrm{h.c.},
\end{aligned}
\label{eq:YukawaLagran}
\end{equation}
where $a,b=1,2,3$,
\begin{equation}
\rho^{ba}_{f}\equiv\dfrac{Y^{2,ba}_{f}}{\cos\beta}-\dfrac{\sqrt{2}\tan\beta\bar{M}_{f}^{ba}}{v},\label{eq:rhos}
\end{equation}
\begin{align}
\Gamma^{hba}_{f} & \equiv\dfrac{\bar{M}_{f}^{ba}}{v}s_{\beta\alpha}+\dfrac{1}{\sqrt{2}}\rho^{ba}_{f}c_{\beta\alpha},\label{eq:Gammafhba}\\
\Gamma^{Hba}_{f} & \equiv\dfrac{\bar{M}_{f}^{ba}}{v}c_{\beta\alpha}-\dfrac{1}{\sqrt{2}}\rho^{ba}_{f}s_{\beta\alpha},\label{eq:GammafHba}\\
\Gamma^{Aba}_{f} & \equiv\begin{cases}
-\dfrac{i}{\sqrt{2}}\rho^{ba}_{f} & \textrm{if }f=u,\\
\dfrac{i}{\sqrt{2}}\rho^{ba}_{f} & \textrm{if }f=d,\ell\,,
\end{cases}\label{eq:GammafAba}
\end{align}
and $s_{\beta\alpha}\equiv\sin(\beta-\alpha)$, $c_{\beta\alpha} \equiv \cos(\beta-\alpha)$.

\subsection{Model parameters}
\label{sec:modelparameters}

Note that the $B$ anomalies are related to second and third generation quarks and leptons. Therefore, we do not consider Yukawa couplings involving the first generation. More specifically, we parametrise the Yukawa matrices as
\begin{equation} 
\rho_{u}=\left(\begin{array}{ccc}
0 & 0 & 0\\
0 & \rho_{u}^{cc} & 0\\
0 & \rho_{u}^{tc} & \rho_{u}^{tt}
\end{array}\right),\qquad\rho_{d}=\left(\begin{array}{ccc}
0 & 0 & 0\\
0 &  0 & 0\\
0 &  0 & \rho_{d}^{bb}
\end{array}\right),\qquad\rho_{\ell}=\left(\begin{array}{ccc}
0 & 0 & \rho_{\ell}^{e\tau}\\
0 & \rho_{\ell}^{\mu\mu} & \rho_{\ell}^{\mu\tau}\\
0 & 0 & \rho_{\ell}^{\tau\tau}
\end{array}\right),\label{eq:Textures}
\end{equation}
where we have extended the pattern of Ref.~\cite{Crivellin:2023sig} to include the second generation diagonal Yukawas for both the lepton and up-type matrices, $\rho_\ell^{\mu\mu}$ and $\rho_u^{cc}$ as well as third generation down-type quark coupling $\rho_d^{bb}$. The diagonal down-type Yukawa coupling $\rho_d^{ss}$ ($\rho_{u}^{ct}$) is however ignored because of strong constraints from the LHC ($b\to s\gamma$), and we choose to consider $\rho_\ell^{\mu\tau,e\tau}$ but not $\rho_\ell^{\tau\mu,\tau e}$ because the simultaneous effect would lead to chirally enhanced effects in $\mu\to e\gamma$ and $\tau\to \mu\gamma,\,e\gamma$~\cite{Crivellin:2013wna,Iguro:2019zlc}.\footnote{The (effective) Yukawa couplings $\rho_f^{ba}$ are derived from the Yukawa couplings $Y_f^{2,ba}$ via Eq.~\eqref{eq:rhos}, with an explicit dependence on $\tan\beta$. Hence, to avoid this explicit dependence on $\tan\beta$ on the model parameters, we will use as fundamental scan parameters the Yukawas $Y_f^{2,ba}$, i.e.~we will be working in the Higgs basis.}

Concerning the scalar potential, we substitute the parameters $\lambda_1-\lambda_5$ by the heavy Higgs masses. Furthermore, the effect of $\lambda_6$ and $\lambda_7$ of EW precision data and flavour observables is automatically included (at the one-loop level). Therefore, the only parameters of the scalar sector are $m_H$, $m_A$, $m_{H^\pm}$, $m_{12}$, $\tan\beta$ and $s_{\beta\alpha}$. 

We used the following ranges for the parameters, based on the findings of Ref.~\cite{Crivellin:2013wna}%
\begin{align}
&m_{12}  \in[-200,\,200]\mathrm{GeV},\qquad m_{H^{\pm}}\in[120,\,140]\mathrm{GeV},\qquad m_{A},\,m_{H}\in[150,\,350]\mathrm{GeV},\nonumber \\
&s_{\beta\alpha} \in[0.98,\,1.0],\qquad\tan\beta\in[0.01,\,10],\qquad Y_{u}^{2,tt}\in[-1.0,\,1.0],\qquad Y_{u}^{2,tc}\in[-0.6,\,0.6],\nonumber \\
 & \mathrm{{Re,\,Im}(}Y_{\ell}^{2,\tau\tau})\in[-0.1,0.1],\qquad Y_{\ell}^{2,e\tau},\,Y_{\ell}^{2,\mu\tau}\in[-0.01,0.01],\nonumber\\ 
&Y_{u}^{2,cc}\in[-0.15,\,0.15],\qquad Y_{d}^{2,bb}\in[-0.2,\,0.2],\qquad Y_{\ell}^{2,\mu\mu}\in[-0.1,\,0.1].
 \label{eq:Ranges}
\end{align}

Note that we allow for a complex $\rho_\ell^{\tau\tau}$ to explain the $b\to c\tau\nu$ anomaly while other couplings are taken to be real. We work close to the alignment limit, i.e.~$s_{\beta\alpha} \in[0.98,\,1.0]$ such that the bounds from SM Higgs signal strength are satisfied.\footnote{We consider here only the bounds from fermionic decays of the Higgs since the di-photon signal strength and always be brought into agreement with the measurement by choosing an appropriate value of $\lambda_7$~\cite{Banik:2024ftv}.}

\section{Observables} \label{sec:Observables}

The flavour-violating couplings of the G2HDM enter into many different processes. Here we present the observables relevant to our analysis and give the corresponding NP contributions.

\subsection{Top decays}

The ATLAS collaboration reported an excess in $t\to bH^+ \to b\bar{b}c$~\cite{ATLAS:2023bzb}
\begin{equation}
    \mathrm{BR}(t\to bH^+ \to b\bar{b}c) = (0.16 \pm 0.06)\%
\end{equation}
for a charged Higgs mass of $m_{H^\pm} = 130$ GeV, which corresponds to a global (local) significance of $2.5\,(3.0)\sigma$. Then the corresponding G2HDM contribution to the decay is given as
\begin{equation}
    \mathrm{BR}(t\to bH^+ \to b\bar{b}c) \approx  \frac{m_t(|\rho_u^{tt}|^2+|\rho_d^{bb}|^2)}{16\pi\Gamma_t}\left(1-\frac{m_{H^\pm}^2}{m_t^2}\right)^2 \frac{3|\rho_u^{tc}|^2 }{3|\rho_u^{tc}|^2+3|\rho_u^{cc}|^2+\sum_{l,l'}|\rho_\ell^{ll'}|^2}\,,
\end{equation}
where $\Gamma_t$ is the total decay width of the top quark. 

For the lepton flavour violating decay $t\to\mu\tau q$~\cite{ATLAS:2024njy}  an upper bound of
\begin{equation}
    \mathrm{BR}(t\to\mu^+\tau^-c) < (8.2 \pm 0.5)\times 10^{-7}.
\end{equation}
at the 90\% CL is found. In the G2HDM the corresponding width is given by 
\begin{equation}
\Gamma(t\to\mu^{+}\tau^{-}c)=\frac{m_{t}^{5}}{3072\,\pi^{3}}\left|c_{lequ}^{1(\mu\tau ct)}\right|^{2},
\end{equation}
with 
\begin{equation}
c_{lequ}^{1(\mu\tau ct)}=\frac{\rho_{u}^{tc*}\rho_{\ell}^{\mu\tau}}{2}\left(\frac{c_{\beta-\alpha}^{2}}{m_{h}^{2}}+\frac{s_{\beta-\alpha}^{2}}{m_{H}^{2}}+\frac{1}{m_{A}^{2}}\right).
\end{equation}
Note that the quark in the final state is only a charm quark given that we are ignoring couplings to first generation quarks.\footnote{ATLAS~\cite{ATLAS:2024oqu} recently searched for $H^+ \to c\bar{s}$ reporting no significant excess. However, in our current setup, we have $\mathrm{BR}(H^{+}\to c\bar{s})<\mathrm{BR}(H^{+}\to c\bar{b})$ and the miss-tagging rate of a strange quark as a $b$ quark is small, such that one can evade this constraint.}

For the decay of a top quark to a charm quark and the SM Higgs, we have in the G2HDM~\cite{Crivellin:2023sig}
\begin{align}
{\rm BR}(t\to hc) & =\frac{m_{t}c_{\beta\alpha}^{2}|\rho_{u}^{tc}|^{2}}{64\pi\Gamma_{t}}\left(1-\frac{m_{h}^{2}}{m_{t}^{2}}\right)^{2} \approx2.4\times10^{-4}\left(\frac{\rho_{u}^{tc}c_{\beta\alpha}}{0.05}\right)^{2},
\end{align}
which can be compared to the current ATLAS~\cite{ATLAS:2023ujo} (CMS~\cite{CMS:2023ufv}) upper limit $\mathrm{BR}(t\to hc)\le4.0\times10^{-4}$ ($3.5\times10^{-4}$).

\subsection{Charged current anomalies in $b\to c\ell\bar\nu$}

The ratios 
\begin{equation}
    R(D^{(*)}) = \frac{{\rm{BR}}(\bar{B}\to D^{(*)}\tau\bar{\nu})}{{\rm{BR}}(\bar{B}\to D^{(*)}l\bar{\nu})},
\end{equation}
with $l = e,\mu$, have been measured by  LHCb~\cite{LHCb:2023uiv,LHCb:2023zxo,LHCb:2024jll}, Belle~\cite{Belle:2015qfa,Belle:2016dyj,Belle:2017ilt,Belle:2019rba}, Belle-II~\cite{Belle-II:2024ami} and BaBar~\cite{BaBar:2012obs,BaBar:2013mob}. The combination provided by HFLAV~\cite{HFLAV2024winter} is
\begin{equation}
 R(D)_{\rm HFLAV} = 0.342 \pm 0.026, \quad R(D^*)_{\rm HFLAV} = 0.287 \pm 0.012\,.
\end{equation}
Using the form factors~\cite{deDivitiis:2007ptj,Kamenik:2008tj} provided in \textsf{SuperIso 4.1}~\cite{Mahmoudi:2007vz,Mahmoudi:2008tp,Mahmoudi:2009zz}, the G2HDM contributions to $R(D)$ and $R(D^{*})$ are given by
\begin{equation}
\begin{aligned}
R(D)\approx\frac{1+1.73\,\mathrm{Re}(g_{S}^{\tau\tau})+1.35\sum\left|g_{S}^{l\tau}\right|^{2}}{3.27+0.57\,\mathrm{Re}(g_{S}^{\mu\mu})+4.8\sum|g_{S}^{l\mu}|^{2}},\,
R(D^{*})\approx\frac{1+0.11\,\mathrm{Re}(g_{P}^{\tau\tau})+0.04\sum\left|g_{P}^{l\tau}\right|^{2}}{4.04+0.08\,\mathrm{Re}(g_{P}^{\mu\mu})+0.25\sum|g_{P}^{l\mu}|^{2}}\label{eq:RDstar},
\end{aligned}
\end{equation}
with $l=e,\,\mu,\,\tau$ and $R(D)_{\rm SM}=0.306$ and $R(D^*)_{\rm SM}=0.247$. The scalar and pseudoscalar couplings $g_{S,P}^{l l'}$ are given in the G2HDM as~\cite{Athron:2021auq},
\begin{eqnarray}
g_{S}^{ll^{\prime}}\equiv\frac{C_{R}^{cb}+C_{L}^{cb}}{C_{SM}^{cb}},\ g_{P}^{ll^{\prime}}\equiv\frac{C_{R}^{cb}-C_{L}^{cb}}{C_{SM}^{cb}},
\end{eqnarray}
where $C_{SM}^{cb}=4G_{F}V_{cb}/\sqrt{2}$ and
\begin{equation}
C_{R}^{cb}=-\frac{(V_{cb}\rho^{bb}_{d}+V_{cs}\rho^{sb}_{d})\rho^{ll^{\prime}*}_{\ell}}{m_{H^{\pm}}^{2}},\quad C_{L}^{cb}=\frac{(V_{tb}\rho^{tc*}_{u}+V_{cb}\rho^{cc*}_{u})\rho^{ll^{\prime}*}_{\ell}}{m_{H^{\pm}}^{2}},
\label{semileptonicWCs}
\end{equation}
which includes the renormalisation group correction factor of $1.5$~\cite{Alonso:2013hga,Jenkins:2013wua,Gonzalez-Alonso:2017iyc,Aebischer:2017gaw} for the Wilson coefficients (WCs).

In addition to $R(D^{(*)})$, the Belle experiment also measured the lepton flavour universality (LFU) ratio $R_{e/\mu}=\mathrm{BR}(\bar{B}\rightarrow D e\bar{\nu})/\mathrm{BR}(\bar{B}\rightarrow D \mu\bar{\nu})$ to be~\cite{Belle:2018ezy}
\begin{equation}
    R_{e/\mu} = 1.01 \pm 0.01 \pm 0.03,
\end{equation}
which can be expressed in the G2HDM as~\cite{Athron:2021auq}
\begin{equation}
R_{e/\mu}\approx\frac{1}{0.9964+\,0.18\,\mathrm{Re} [g_{S}^{\mu\mu}]+1.46\sum_l\left|g_{S}^{l\mu}\right|^{2}}\,,
\end{equation}
where we have obtained the NP leptonic contributions by integrating the heavy quark effective theory amplitudes of the scalar type operators from Refs.~\cite{Murgui:2019czp, Tanaka:2012nw}.

\subsection{Leptonic meson decays}

The fully leptonic decays of mesons can receive chirally enhanced effects from (pseudo-) scalar currents. The total decay width in the G2HDM is~\cite{HernandezSanchez:2012eg,Jung:2010ik,Iguro:2017ysu}
\begin{eqnarray}
\mathrm{BR}(M_{ij}\to \ell\nu)=G_{F}^{2}m_{l}^{2}f_{M}^{2}\tau_{M}|V_{ij}|^{2}\frac{m_{M}}{8\pi}\left(1-\frac{m_{\ell}^{2}}{m_{M}^{2}}\right)^{2}\left[|1-\Delta_{ij}^{\ell\ell}|^{2}+|\Delta_{ij}^{\ell^{\prime}\ell}|^{2}\right],
\end{eqnarray}
where $i$, $j$ are the valence quarks of the meson $M$, $f_{M}$ is
its decay constant and $\Delta_{ij}^{ll^{\prime}}$ is the NP
correction given by 
\begin{eqnarray}
\Delta_{ij}^{\ell^{\prime}l}=\bigg(\frac{m_{M}}{m_{H^{\pm}}}\bigg)^{2}Z_{\ell^{\prime}l}^{*}\bigg(\frac{Y_{ij}m_{u_{i}}+X_{ij}m_{d_{j}}}{V_{ij}(m_{u_{i}}+m_{d_{j}})}\bigg),\quad\,\ell\neq \ell^\prime,\,\,\,\,\ell,\ell^{\prime}=2,3\,,
\end{eqnarray}
with
\begin{equation}
X_{ij}=\frac{v}{\sqrt{2}m_{d_{j}}}V_{ik}\,\rho^{kj}_{d},\qquad Y_{ij}=\frac{v}{\sqrt{2}m_{u_{i}}}\,\rho^{ki*}_{u}\,V_{kj},\qquad Z_{\ell^\prime \ell}=\frac{v}{\sqrt{2}m_{j}}\,\rho^{\ell^\prime \ell}_{\ell}.
\end{equation}

In particular, we consider $\mathrm{BR}(D_{s}\to \mu\bar\nu)=(5.43\pm0.15)\times 10^{-3}$, $\mathrm{BR}(D_{s}\to \tau\bar\nu)=(5.32\pm0.11)\times 10^{-2}$ and $\mathrm{BR}(B_c\to\tau\bar\nu)$. Regarding the latter, the theoretical prediction within the SM is still unclear and upper limits of $60\%$~\cite{Alonso:2016oyd,Blanke:2018yud,Aebischer:2021ilm} are still possible. To be conservative, we define a likelihood function allowing values for $\mathrm{BR}(B_c\to\tau\bar\nu)\leq70\%$.
The expression for the SM plus the G2HDM contribution for ${\rm {BR}}(B_{c}\to\tau\bar\nu)$ is given at tree level \cite{Iguro:2022yzr} as,
\begin{equation}
\mathrm{BR}(B_c\to\tau\bar\nu)\approx\mathrm{BR}(B_c\to\tau\bar\nu)_{\mathrm{SM}}\,\left[ \left|1-4.35\,C^{\tau\tau}_{L} \right|^2 +4.35^2\left(\left|C^{e\tau}_{L} \right|^2 +\left|C^{\mu\tau}_{L} \right|^2\right) \right]. 
\end{equation}

\subsection{Neutral current anomalies: $b\to s$ transitions}
\label{sec:btos}
Global fits to $b\to s \ell^+ \ell^-$ observables favour $C_9^U\approx -1$ at the $5\sigma$ level~\cite{Buras:2022qip,Neshatpour:2022pvg,Gubernari:2022hxn,Ciuchini:2022wbq,Alguero:2023jeh,Wen:2023pfq,Capdevila:2023yhq,Athron:2023hmz}. The most relevant observables include $P_5^\prime$~\cite{Descotes-Genon:2012isb,LHCb:2015svh,LHCb:2020lmf,LHCb:2020gog}\footnote{Recently, the CMS collaboration~\cite{CMS:2024tbn} made competitive measurements of the angular observables in good agreement with LHCb data confirming the anomaly. For more details and complete expressions for the angular observables the reader is referred to Refs.\,\cite{Altmannshofer:2008dz,Bhom:2020lmk,Athron:2021auq}.}, the total branching ratio and angular observables in $B_s\to\phi\,\mu^+\mu^-$~\cite{LHCb:2015wdu,LHCb:2021zwz,LHCb:2021xxq} as well as the BR$(B\to K\mu^+\mu^-)$~\cite{LHCb:2014cxe,LHCb:2016ykl,Parrott:2022zte}, which are fully compatible with semi-inclusive observables~\cite{Isidori:2023unk}.  This inspires lepton flavour universal NP models with vectorial couplings to leptons and left-handed couplings to bottom and strange quarks that may relax the tension.

In the G2HDM model, mainly the charm loop contributes to $C_9^U$ via an off-shell photon penguin~\cite{Jager:2017gal,Bobeth:2014rda,Iguro:2018qzf,Crivellin:2019dun,Kumar:2022rcf,Iguro:2023jju,Crivellin:2023saq} and we obtain~\cite{Crivellin:2019dun},
\begin{align}
    \Delta C_9^U(\mu_b)\approx &-0.52\left(\frac{|\rho_u^{tc}|^2-|\rho_u^{cc}|^2}{0.5^2}\right)+0.50\left(\frac{\rho_u^{tc*}\rho_u^{cc}}{0.01}\right).
    \label{eq:C9U}
\end{align}
We see that a sizable coupling $\rho_u^{tc}$ is necessary if $\rho_u^{cc}\approx0$ is assumed while the product $\rho_u^{tc*}\rho_u^{cc}$ has a CKM enhancement w.r.t.~the SM. While in the previous work~\cite{Crivellin:2023sig} the value of $\rho_u^{cc}$ was set to zero, it could play an important role in obtaining values in agreement with model-independent fits.
In addition to the contributions to $C_9^U$, the closely-related quark level decay $b\to c\bar{c}s$ can noticeably affect the $B_s$ lifetime~\cite{Jager:2017gal,Jager:2019bgk} and potentially constrain the G2HDM.
Since it is not easy to control the exclusive decay $b\to c\bar{c}s$, the lifetime ratio $\tau_{B_s}/\tau_{B_d}$ is typically used, which can be calculated with a heavy bottom quark expansion. Nevertheless, in our scenario thanks to the CKM suppressed $b\to c\bar{c}d$ interaction this shift in the ratio is cancelled, significantly relaxing the constraint and hence we will not consider it henceforth.

Regarding scalar operators with coefficients $C_{Q_{1,2}}$, the most sensitive observable is the branching ratio $\mathrm{BR}(B_{s}\rightarrow\mu^{+}\mu^{-})$ which also depends on $C_{10}^{(\prime)}=C_{10}^{(\prime)\mathrm{SM}}+\Delta C_{10}^{(\prime)}$~\cite{Bhom:2020lmk}:
\begin{align}
\mathrm{BR}(B_{s} & \rightarrow\mu^{+}\mu^{-})=\dfrac{G_{F}^{2}\alpha^{2}}{64\pi^{3}}f_{B_{s}}^{2}\tau_{B_{s}}m_{B_{s}}^{3}\big|V_{tb}V_{ts}^{*}\big|^{2}\sqrt{1-\frac{4m_{\mu}^{2}}{m_{B_{s}}^{2}}}\nonumber\\
 & \times\left[\left(1-\frac{4m_{\mu}^{2}}{m_{B_{s}}^{2}}\right)\left|\dfrac{m_{B_{s}}\left(C_{Q_{1}}- C_{Q_{1}}^{'}\right)}{(m_{b}+m_{s})}\right|^{2} 
 +\left|\dfrac{m_{B_{s}}\left(C_{Q_{2}}-C_{Q_{2}}^{'}\right)}{\left(m_{b}+m_{s}\right)}-2\left(C_{10}-C_{10}^{\prime}\right)\frac{m_{\mu}}{m_{B_{s}}}\right|^{2}\right],
\label{eq:Bsmumu}
\end{align}
where $f_{B_{s}}$ is the $B_s$ meson decay constant, $\tau_{B_{s}}$ is its
mean lifetime and the WCs are given in Refs.\,\cite{Crivellin:2019dun,Athron:2021auq}. The experimental data for all these observables are taken from the \textsf{HEPLike} package~\cite{Bhom:2020bfe}, whereas the theoretical predictions are extracted from \textsf{SuperIso}~\cite{Mahmoudi:2007vz, Mahmoudi:2008tp, Mahmoudi:2009zz}.

For the radiative decay $\bar{B}\rightarrow X_{s}\gamma$, the contributions from the G2HDM are taken from Refs.~\cite{Czarnecki:1998tn,Misiak:2006zs,Misiak:2006ab,Czakon:2015exa,Misiak:2017bgg,Misiak:2020vlo} as implemented in \textsf{SuperIso}~\cite{Mahmoudi:2007vz, Mahmoudi:2008tp, Mahmoudi:2009zz}. As a function of the $C_7$ and $C'_7$ Wilson coefficients, the $b\to s\gamma$ transition rate can be written as
\begin{equation}
\Gamma(b\rightarrow s\gamma)=\frac{G_{F}^{2}}{32\pi^{4}}\big|V_{tb}V_{ts}^{*}\big|^{2}\alpha_{{\rm em}}\,m_{b}^{5}\,\left(\vert C_{7{\rm eff}}\,(\mu_{b})\vert^{2}+\vert C_{7{\rm eff}}^{\prime}(\mu_{b})\vert^{2}\right),
\end{equation}
where $C^{(')}_{7\rm eff} = C^{(')\rm SM}_{7\rm eff} + \Delta C^{(')}_{7\rm eff}$ is the effective Wilson coefficient (see below), and the experimental measurement is $\mathrm{BR}(B\rightarrow X_{s}\gamma)\times10^{4}=3.49\pm0.19$~\cite{HFLAV:2022esi}. Taking into account the $1/\sqrt{2}$ factor difference in the notations with respect to Ref.\,\cite{Crivellin:2019dun}, the dominant top quark contribution for the $\Delta C_7$ WC is given by
\begin{align}
\Delta{C_7}^{t,\,H^\pm} = - \dfrac{1}{{36}}\dfrac{m_W^2}{m_{H^\pm}^2}\dfrac{{V_{ks}^*\rho^{kt}_u\rho^{nt*}_{u}{V_{nb}}}}{{g_2^2{V_{tb}}V_{ts}^*}}f_{1}\left(\frac{m_t^2}{m_{H^\pm}^2}\right),
\label{bsgamma1}
\end{align}
 where $f_{1}(x)$ is a loop function one can find in Ref.\,\cite{Crivellin:2019dun} from where we use all expressions for computing the WCs. While the charm quark contribution is obtained as,
\begin{align}
\Delta{C_7}^{c,\,H^\pm}(\mu) = - \dfrac{7}{{36}}\dfrac{m_W^2}{m_{H^\pm}^2}\dfrac{{V_{ks}^*\rho^{kc}_u\rho^{nc*}_{u}{V_{nb}}}}{{g_2^2{V_{tb}}V_{ts}^*}}.
\label{C8light}
\end{align}
Numerically with $m_{H^\pm}= 130$\,GeV one gets
\begin{equation}
\Delta C_{7\rm eff}(\mu_b)\approx-0.174\left(|\rho_u^{tc}|\right)^2-0.046\left(|\rho_u^{tt}|\right)^2,\label{eq:C7}   
\end{equation}
which can be used for understanding possible explanations of the anomaly in $t \rightarrow bH^+ \rightarrow b\bar{b}c$.  
We note that $C_7$ of the SM is negative which fixes the convention and hence $\Delta C_{7\rm eff}$ constructively interferes with the SM contribution. Furthermore, we have implemented the respective RGE effects in \gambit including the mixing with the $C_8$ WC. From model independent fits, we require the NP contribution to be within $-0.04\le\Delta C_{7}(\mu_b)\le0.04$ \cite{Alguero:2023jeh}. 
Additionally, once we turn on the second generation diagonal quark Yukawas we will have mixed terms given by the semi-numerical expression
\begin{align}
\Delta C_{7\rm eff}^{\mathrm{mix}}(\mu_{b}) & \approx0.105\left(\frac{\rho_{u}^{tc*}\rho_{u}^{cc}}{0.025}\right)-0.028\left(\frac{\rho_{u}^{cc}\rho_{d}^{bb}}{0.025}\right)+0.19\left(\frac{\rho_{u}^{tt*}\rho_{d}^{bb}}{0.025}\right).\label{eq:C7-1}
\end{align}
We note that these terms can take both signs depending on the sign of the coupling product and hence potentially cancel the negative contribution from Eq.\,(\ref{eq:C7}).

Last, the recent measurement of $\mathrm{BR}(B^{+}\to K^{+}\nu\bar\nu)$ by Belle II is yet another exciting hint of NP~\cite{Belle-II:2023esi}. Given its close relation to the $b\to s\ell^{+}\ell^{-}$ decays one could expect both a $Z$ penguin and box diagram contribution enhancement in the G2HDM. However, the necessary WCs turn out to be very suppressed by all other flavour constraints at the $2\,\sigma$ level, implying an SM-like prediction for $\mathrm{BR}(B^{+}\to K^{+}\nu\bar\nu)$.

\subsection{$B_s-\overline{B_s}$ mixing}

The mass difference of the $B_s$ and $\overline{B_s}$ mesons in the presence of NP is
\begin{equation}
\Delta M_{B_{s}}^{\mathrm{G2HDM}}=\Delta M_{B_{s}}^{\mathrm{SM}}\left(1+\frac{M_{12}^{\mathrm{NP}}}{M_{12}^{\mathrm{SM}}}\right),\label{eq:DeltaMs_no_running}
\end{equation}
where $\Delta M_{B_{s}}^{\mathrm{SM}}=(18.2^{+0.6}_{-0.8})\,\mathrm{ps}^{-1}$ is the SM prediction from Ref.\,\cite{Crivellin:2023saq} and 
\begin{align}
M_{12}^{\mathrm{SM}} & =\frac{G_{F}^{2}\,m_{W}^{2}M_{B_{s}}f_{B_{s}}^{2}\mathcal{B}_{B_{s}}\eta_{B}}{12\pi^{2}}\left(V_{tb}V_{ts}^{*}\right)^{2}S_{0},\label{eq:M12_SM}
\end{align}
where the parameter $\mathcal{B}_{B_{s}}\approx0.841$ is the so-called \textit{bag factor}, $\eta_{B}=0.8393\pm0.0034$ is due to QCD corrections and $S_{0}\approx2.35$ is an Inami-Lim function \cite{Lenz:2010gu} including electroweak corrections.

\begin{figure}[H]
\begin{centering}
\includegraphics[scale=0.30]{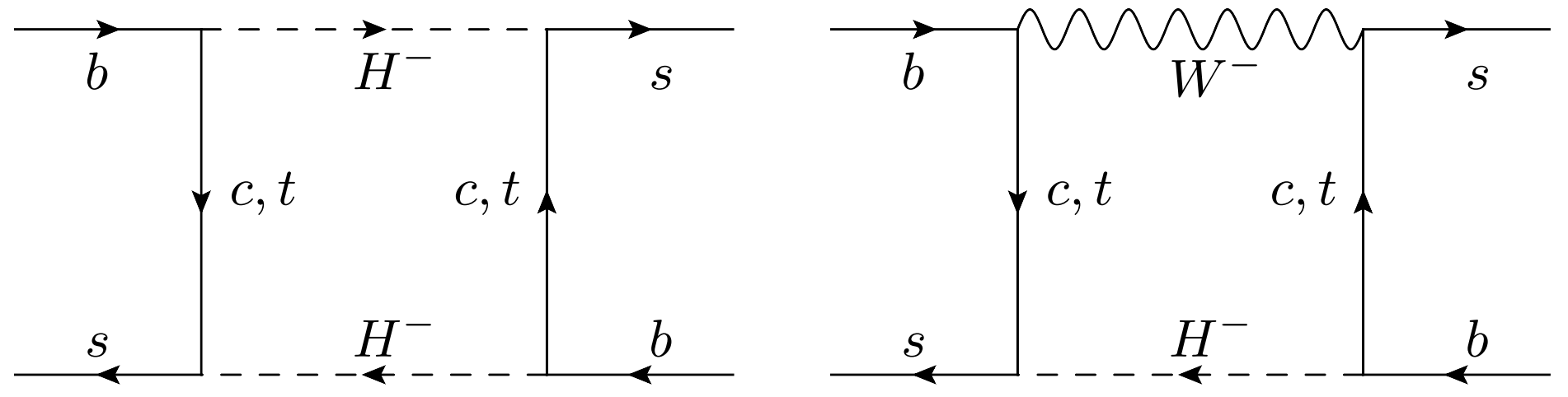}
\par\end{centering}
\caption{\emph{Box diagrams relevant for $B_s-\overline{B_s}$ mixing.}\label{fig:box_diagrams}}
\end{figure}

Using both the theoretical expressions of Refs.\,\cite{Iguro:2017ysu,Crivellin:2019dun} and doing an independent calculation with \texttt{FeynCalc} \cite{Shtabovenko:2016sxi} with the model files provided by \texttt{FeynRules} \cite{Degrande:2014vpa} as a cross check, we compute the G2HDM contribution to the mass difference at one loop level. For simplicity, we show here the resultant expressions evaluated at the charged Higgs mass $m_{H^{\pm}}=130\,\mathrm{GeV}$, even though later we use the full expression depending on $m_{H^{\pm}}$ in the scans. We find for 
\begin{align}
\Delta M_{B_s}^{\mathrm{G2HDM}}/\Delta M_{B_s}^{\mathrm{SM}}\approx&-0.0055|\rho_{u}^{tc}|^{2}+1.73|\rho_{u}^{tt}|^{2}-0.9\,\rho_{u}^{tc*}\rho_{u}^{tt}\nonumber \\
&+87\rho_{u}^{tc}\rho_{u}^{cc}(|\rho_{u}^{cc}|^{2}-|\rho_{u}^{tc}|^{2})\text{}+1046|\rho_{u}^{cc}|^{2}|\rho_{u}^{tc}|^{2}\,,\label{eq:Bsmix_repr}
\end{align}
where the first and second lines are the quadratic and quartic terms related to the $W$--$H^-$ and $H^-$--$H^-$ box diagrams in Fig.\,\ref{fig:box_diagrams}, respectively.\footnote{Regarding the width difference $\Delta\Gamma_{s}$ of the $B_{s}-\overline{B_s}$ system, we corroborated that the contribution of NP to the CP violating phase $\phi^{\Delta}_{s}$ in Ref.\,\cite{Lenz:2010gu} is negligible given that we do not consider complex Yukawa couplings except for $\rho_\ell^{\tau\tau}$. }

\subsection{Lepton flavour (universality) violation}

The ratio
\begin{eqnarray}
\left(\frac{g_{\mu}}{g_{e}}\right)^{2}=\frac{\mathrm{BR}(\tau\to\mu\bar{\nu}\nu)}{\mathrm{BR}(\tau\to e\bar{\nu}\nu)}\frac{f(m_{e}^{2}/m_{\tau}^{2})}{f(m_{\mu}^{2}/m_{\tau}^{2})}\approx1+\sum_{i,j=\mu,\tau}\left(0.25\,|R_{ij}|^{2}-0.11\,\mathrm{Re}(R_{ij})\right),
\end{eqnarray}
where $f(x)=1-8x+8x^{3}-x^{4}-12x^{2}\,\log x$ 
and $R_{ij}$ is the BSM scalar contribution for the test of lepton flavour universality in the tau sector. In the G2HDM at tree level\footnote{We confirmed that the dominant contributions coming from one-loop diagrams \cite{Krawczyk:2004na,Abe:2015oca, Crivellin:2015hha} are negligible for $|\rho_\ell^{\tau\tau}|<0.2$.} we have
\begin{eqnarray}
R_{ij}=\frac{\upsilon^{2}}{2\,m_{H^{\pm}}^{2}}\,\rho^{\tau i}_{\ell}\,\rho^{j\mu\,*}_{\ell}.\label{R scalar}
\end{eqnarray}

The corresponding experimental measurement is taken from the latest Belle II, $|g_{\mu}/g_{e}|=0.9974\pm0.0019$~\cite{Corona:2024nnn}.

In the G2HDM lepton flavour violating decays of $\tau$ and $\mu$ leptons are induced by one-loop diagrams and enhanced 2-loop contributions known as Barr-Zee~\cite{Barr:1990vd,Abe:2013qla,Ilisie:2015tra}. Here we use the expressions from Ref.\,\cite{Athron:2021auq} for the predictions of  $\tau\rightarrow\mu\gamma$, $\tau\to3\mu$ and $\mu\to e\gamma$. They are compared to the upper experimental limits from the PDG~\cite{ParticleDataGroup:2024cfk} and the MEG II collaboration \cite{MEGII:2023ltw}.

For the Higgs LFV decays we follow Ref.\,\cite{Crivellin:2023sig} and use
\begin{align}
{\rm {BR}}(h\to l\tau) & =\frac{c_{\beta\alpha}^{2}m_{h}}{16\pi^{2}\Gamma_{h}}\left(|\rho_{\ell}^{l\tau}|^{2}+|\rho_{\ell}^{\tau l}|^{2}\right),\label{eq:t_bbc}
\end{align}
where $l=e,\,\mu$,
with experimental bounds provided by CMS and ATLAS~\cite{CMS:2021rsq, ATLAS:2023mvd}.  Note this data also contains anomalies, but here we do not consider these and instead simply apply them as the upper limits presented in Refs.\ \cite{CMS:2021rsq, ATLAS:2023mvd}.

\subsection{Higgs searches at colliders}

The relative coupling strength $\kappa_\tau$ for $h\tau\bar{\tau}$ is defined as the ratio $\kappa_\tau^2\equiv \Gamma_\tau/\Gamma^{\mathrm{SM}}_\tau$ where $\Gamma_\tau$ is the partial decay width into a pair of taus, and measured to be $\kappa_\tau = (0.93\pm0.07)$ by ATLAS~\cite{ATLAS:2022vkf} and $\kappa_\tau = (0.92 \pm 0.08)$ by CMS~\cite{CMS:2022dwd}. It is affected in the G2HDM as
\begin{align}
    \kappa_\tau= \biggl{|}s_{\beta\alpha}+\frac{ \rho_\ell^{\tau\tau} c_{\beta\alpha} }{\frac{\sqrt{2}m_\tau}{v}} \biggl{|}.
\end{align}
Concerning direct search for the new Higgs bosons, $H$, $A$ and $H^\pm$, we use the exclusion limits computed by \textsf{HiggsBounds}~\cite{Bechtle:2008jh, Bechtle:2011sb, Bechtle:2013wla, Bechtle:2015pma} and \textsf{HiggsSignals}~\cite{Bechtle:2013xfa, Stal:2013hwa}. In addition, we add the limits from searches for heavy Higgses with flavour-violating couplings, not presently included in \textsf{HiggsBounds} or \textsf{HiggsSignals}. Here, the ATLAS search for the production of a heavy Higgs decaying via flavour-violating couplings resulting in a pair of same-sign tops can set a strong lower limit on the masses of heavy Higgses~\cite{ATLAS:2023tlp}. Nevertheless, this limit is not very effective in our study since we have no mixing between the first and third generation, $\rho_u^{tu}=0$, and the off-diagonal Yukawas are small, $\rho_u^{tc}<0.5$. The CMS search for the production of two heavy Higgses from off-shell $Z$, decaying to four taus is said to exclude the possibility of fitting the anomalous magnetic moment of the muon (see below) in the lepton-specific 2HDM~\cite{CMS:2023mkp}.\footnote{Moreover, the reinterpretation of neutralino search resulting in multi-taus and missing energy by ATLAS \cite{ATLAS:2022nrb} is said to exclude the scenario too \cite{Iguro:2023tbk}.} The limits provided on the cross sections and branching ratios can be re-interpreted in a model-agnostic way, and thus applied to our scenario.  However, departing from the limit of the lepton-specific 2HDM, the di-Higgs production from an off-shell $Z$ is generically small, such that the effect of the constraint is weakened. Moreover, in our scenario, due to the presence of additional quark Yukawa couplings, the tauonic branching ratio is diluted.

\subsection{Oblique parameters and $m_W$ mass}
\label{sec:stu}

The oblique parameters $S$, $T$ and $U$~\cite{Peskin:1990zt,Peskin:1991sw} parameterise NP correction to the electroweak gauge boson propagators. In particular, they are sensitive to the $W$ mass if the EW sector of the SM is fixed via $G_F$, $\alpha$ and the $Z$ mass. However, the situation for the $W$ mass measurement is not clear at the moment. While the CDF-II collaboration found a $m_W$ value~\cite{CDF:2022hxs} which is $7\sigma$ above the SM prediction, the measurements from LEP \cite{ALEPH:2013dgf}, D0 \cite{D0:2012kms} and the LHC~\cite{LHCb:2021bjt,ATLAS:2024erm} included in the PDG average~\cite{ParticleDataGroup:2024cfk} are in better agreement with the SM.\ Furthermore, very recently the CMS collaboration reported a preliminary result of $m_W$ consistent with the EW global fit of the SM~\cite{CMSWmass}, with a precision comparable with the one from CDF-II.

There are significant tensions between the LHC and the CDF-II measurements, resulting in a poor compatibility when all the measurements are combined~\cite{LHC-TeVMWWorkingGroup:2023zkn}. Out of all possible combinations where a single measurement is removed, the best compatibility is obtained without the CDF-II measurement~\cite{LHC-TeVMWWorkingGroup:2023zkn}.
Therefore, we consider as the default option that in which CDF-II is disregarded, leading to
\begin{equation}
\begin{aligned} \mathrm{PDG:}\,\,\,S=-0.04\pm0.10,\,\,\,T=0.01\pm0.12,\,\,\,U=-0.01\pm0.09,\,\label{eq:STUPDG2024}
\end{aligned}
\end{equation}
with correlation matrix $\rho$ given by,
\begin{equation}
\rho^{\mathrm{PDG}}=\left(\begin{array}{ccc}
1 & 0.93 & -0.70\\
0.93 & 1 & -0.87\\
-0.70 & -0.87 & 1
\end{array}\right).
\end{equation}
However, we also compare our results to the alternative case in which only the CDF-II measurement is used, resulting in~\cite{Lu:2022bgw}
\begin{equation}
\begin{aligned} \mathrm{CDF:}\,\,\,S=0.06\pm0.10,\,\,\,T=0.11\pm0.12,\,\,\,U=0.14\pm0.09,\,\end{aligned}
\end{equation}
with correlations
\begin{equation}
\rho^{\mathrm{CDF}}=\left(\begin{array}{ccc}
1 & 0.90 & -0.59\\
0.90 & 1 & -0.85\\
-0.59 & -0.85 & 1
\end{array}\right).
\end{equation}
Here we use the \textsf{2HDMC 1.8} package~\cite{Eriksson:2009ws} to include these constraints where general expressions for the $S$, $T$ and $U$ parameters from Refs.\,\cite{Grimus:2007if,Grimus:2008nb} are used. Unless otherwise explicitly stated, we use the PDG values for the $S$, $T$ and $U$ parameters from Eq.~\eqref{eq:STUPDG2024}.

\subsection{Anomalous magnetic moment of the muon: $(g-2)_\mu$}

While the situation for the direct experimental measurement of $(g-2)_\mu$ is clear~\cite{Muong-2:2023cdq,PhysRevLett.126.141801} \begin{eqnarray}
a_{\mu}^{\textrm{Exp}} = (11659205.9\pm2.2)\times10^{-10}\,,
\end{eqnarray}
the SM prediction is puzzling. The SM value of $a_{\mu}^{\textrm{WP}} = (11659181.0\pm4.3)\times10^{-10}$ computed by the $g-2$ Theory Initiative's White Paper 
(WP)~\cite{Aoyama:2020ynm}, which is based on work from \cite{davier:2017zfy,keshavarzi:2018mgv,colangelo:2018mtw,hoferichter:2019gzf,davier:2019can,keshavarzi:2019abf,kurz:2014wya,melnikov:2003xd,masjuan:2017tvw,Colangelo:2017fiz,hoferichter:2018kwz,gerardin:2019vio,bijnens:2019ghy,colangelo:2019uex,colangelo:2014qya,Blum:2019ugy,aoyama:2012wk,Aoyama:2019ryr,czarnecki:2002nt,gnendiger:2013pva}, gives a deviation,
\begin{equation} \label{eqn:gm2discrepancyWP}
    \Delta a_{\mu}^\textrm{WP} = (24.9\pm4.8)\times10^{-10}.
\end{equation}
which is a $5.1\sigma$ tension with the direct measurement. However, that prediction does not include the BMW lattice calculation for the Hadronic Vacuum Polarisation (HVP) contribution~\cite{Borsanyi:2020mff,Boccaletti:2024guq}.  Replacing the HVP contribution from the WP with the latest BMW calculation results in 
\begin{equation} \label{eqn:gm2discrepancyBMW}
    \Delta a_{\mu}^\textrm{BMW} = (4.0\pm4.4)\times10^{-10},
\end{equation}
The BMW calculation has been confirmed by other lattice groups~\cite{Ce:2022kxy,ExtendedTwistedMass:2022jpw,RBC:2023pvn,FermilabLatticeHPQCD:2023jof} in an intermediate Euclidean time window~\cite{RBC:2018dos}.  Note also that the recent  CMD-3 \cite{CMD-3:2023alj} measurement of $e^+e^-\to $hadrons, gives systematically larger cross-section than BaBar~\cite{BaBar:2012bdw} and KLOE~\cite{KLOE-2:2017fda}.
Finally, $\tau$-data driven results are also consistent with the lattice calculations~\cite{Masjuan:2023qsp,Davier:2023fpl}. Therefore, we will use the BMW value for the fit in the main text but show the impact if a large NP effect from the G2HDM in $(g-2)_\mu$ was preferred in the appendices, namely we show in appendix~\ref{sec:BZ-G2HDM} the different contributions from the model and in appendix~\ref{sec:wp} the results obtained if using the WP value.

\section{Results}\label{sec:Results}

We now perform a global fit of the G2HDM to the observables discussed in Sec.~\ref{sec:Observables}, using the \gambit~\cite{Athron:2017ard,grev} framework. For this, we extend the \textsf{FlavBit}~\cite{Workgroup:2017myk}, \textsf{PrecisionBit}~\cite{GAMBITModelsWorkgroup:2017ilg} and \textsf{ColliderBit}~\cite{GAMBIT:2017qxg} modules of~\textsf{GAMBIT} to compute the observables and likelihoods for the G2HDM as described in the previous section. We also make use of various external codes: \textsf{SuperIso 4.1}~\cite{Mahmoudi:2007vz,Mahmoudi:2008tp,Mahmoudi:2009zz,Neshatpour:2021nbn} for computing flavour observables, \textsf{2HDMC 1.8}~\cite{Eriksson:2009ws} for precision electroweak constraints, \textsf{HEPLike}~\cite{Bhom:2020bfe} for likelihoods of $b\to s\ell^+\ell^-$ observables. We employ the differential evolution sampler \textsf{Diver 1.0.4}~\cite{Workgroup:2017htr} to explore the parameter space\footnote{The results of all the scans, ran in the LUMI supercluster in Kajaani, Finland, accumulated a total of 60 million parameter samples from 10 independent scans, using approximately 2400 CPU hours.} and the plotting script \textsf{pippi}~\cite{Scott:2012qh} to produce all the figures below. To validate the implementation of the observables, we confirmed that the predictions for them calculated by \gambit (see Table~\ref{tab:observables}) agree with Ref.~\cite{Crivellin:2023sig} within the corresponding theoretical and parametric errors for one of the benchmark scenarios presented there (henceforth called ``BM3''). Some small differences are due to different choices of the SM predictions vs experimental input. 

To provide more information on the impact of different observables and parameters we perform multiple fits.  First, we consider only the parameters used in Ref.\ \cite{Crivellin:2023sig} with a limited set of observables to examine how the charged-current $B$ anomalies can be explained and to study the implications for $b\to s\ell^+\ell^-$ data.  We then perform a more comprehensive fit, including additional observables and increasing the number of Yukawa parameters. In our main analysis we treat $(g-2)_\mu$ and the $W$ mass as {\it constraints} on new physics, using the BMW prediction for HVP contributions \cite{Boccaletti:2024guq} for the former, and the oblique parameters~\cite{ParticleDataGroup:2024cfk} from fits that exclude the CDF-II measurement for the latter.  However we also compare this to another fit showing the impact of using the CDF-II measurement instead.  Similarly, for the purpose of comparison, in the appendices we show how results change if we instead use the the HVP contributions given in the White Paper, which implies a preference for a significant BSM contribution.   

\begin{figure}[h]
\begin{centering}
\includegraphics[scale=0.6]{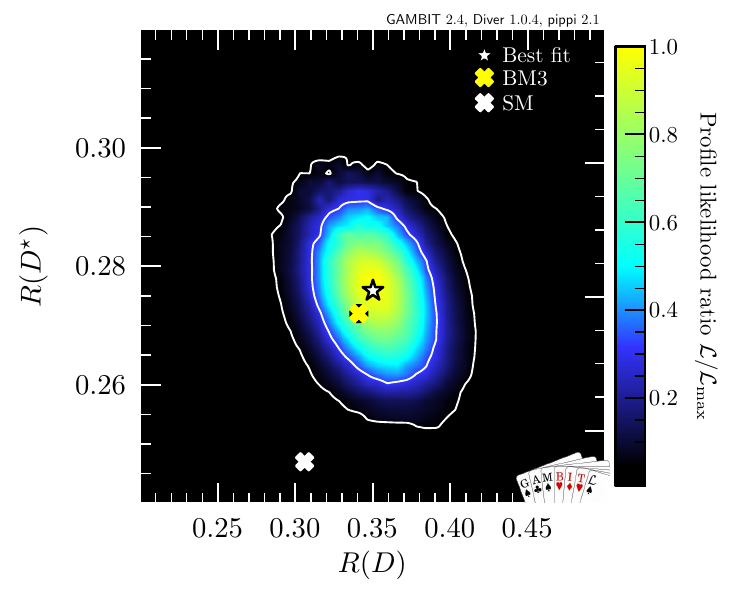}$\quad$\includegraphics[scale=0.6]{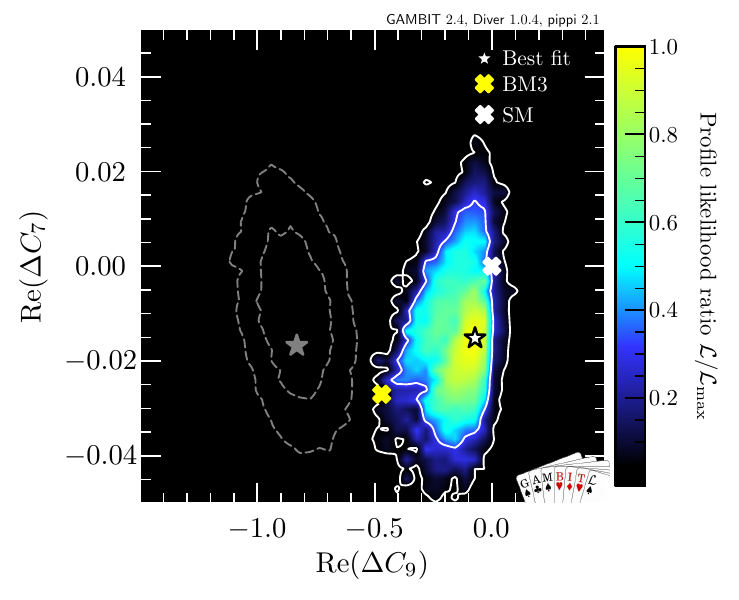}
\par\end{centering}
\caption{\emph{Profile likelihood ratios for $R(D)$ and $R(D^*)$ (left) and $\Delta C_9$ and $\Delta C_7$ (right). The white star denotes the best-fit point, the white cross the SM prediction, and the yellow cross corresponds to BM3 from Ref.~\cite{Crivellin:2023sig}. White contours around the best-fit point are the $1\sigma$ and $2\sigma$ confidence intervals (calculated with two degrees of freedom).  The grey contours on the right shows the region preferred by the model-independent fit to $b\to s\ell^{+}\ell^{-}$ transitions.}\label{fig:Results_BM3}}
\end{figure}

In our first scan we consider the reduced parameter space of $\{m_{12}$, $m_{H^{\pm}}$, $m_{A}$, $m_{H}$, $s_{\beta\alpha}$, $\tan\beta$, $Y_{u}^{2,tt}$, $Y_{u}^{2,tc}$, $Y_{\ell}^{2,\tau\tau}$, $Y_{\ell}^{2,\mu\tau}$, $Y_{\ell}^{2,e\tau}\}$ which matches the parameters considered in Ref.~\cite{Crivellin:2023sig}. For this, we exclude $b\to s\ell^+\ell^-$ from the fit and predict the contribution to the WCs of the relevant effective operators i.e.~$\Delta C_{9,7}$. The resulting profile likelihood ratio is shown in Fig.~\ref{fig:Results_BM3}, where the best-fit point is marked with a white star and the white contours are the boundaries of the $1\sigma$ and $2\sigma$ regions. Also the benchmark point 3 (BM3) of Ref.~\cite{Crivellin:2023sig} is depicted in yellow and the SM prediction in white. The panel on the left shows that a good fit to $R(D^*)$ can be obtained. The panel on the right illustrates that despite disregarding $b\to s\ell^+\ell^-$ data, the fit has a preference for negative values of $\Delta C_9$, in agreement with the model-independent fit (grey contours)~\cite{Greljo:2022jac,Alguero:2023jeh}.  However, the magnitude of $\Delta C_9$ is significantly smaller within this setup compared to the one obtained from the model-independent fit. Hence, we will consider if extending the set of parameters can resolve this tension.

\begin{figure}[h]
\begin{centering}
\includegraphics[scale=0.6]{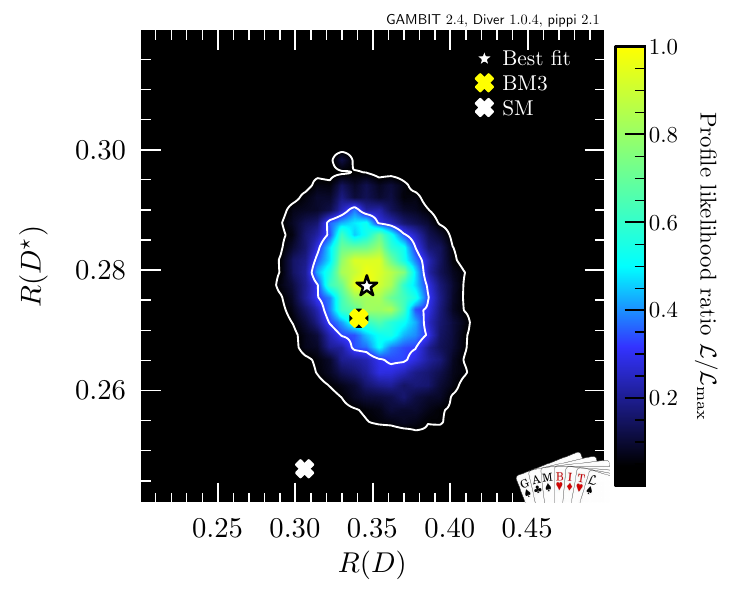}$\quad$\includegraphics[scale=0.6]{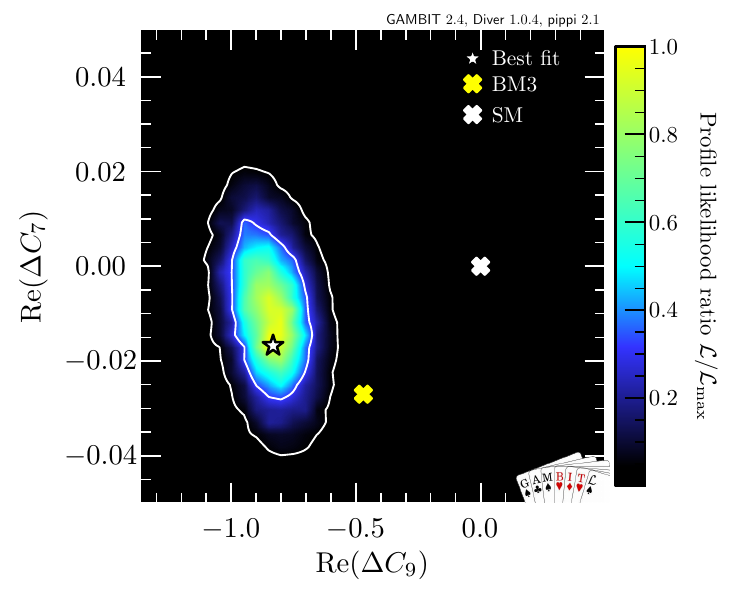}
\includegraphics[scale=0.6]
{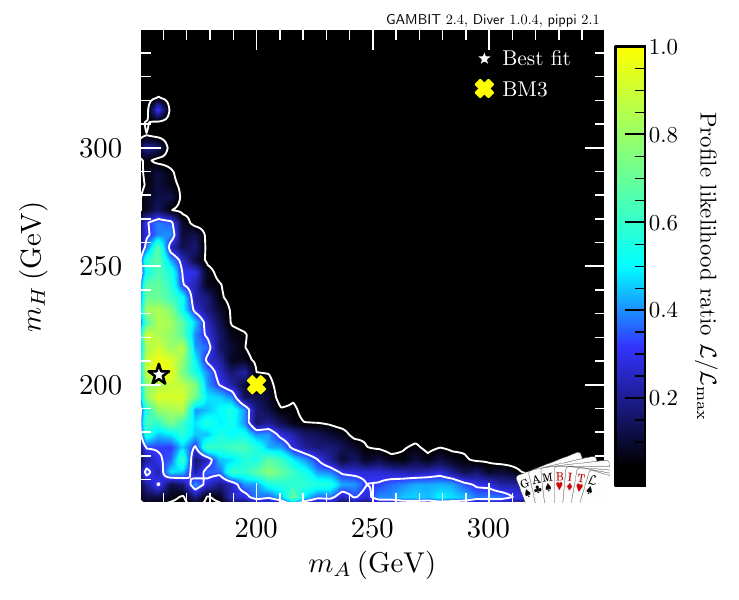}
\includegraphics[scale=0.6]{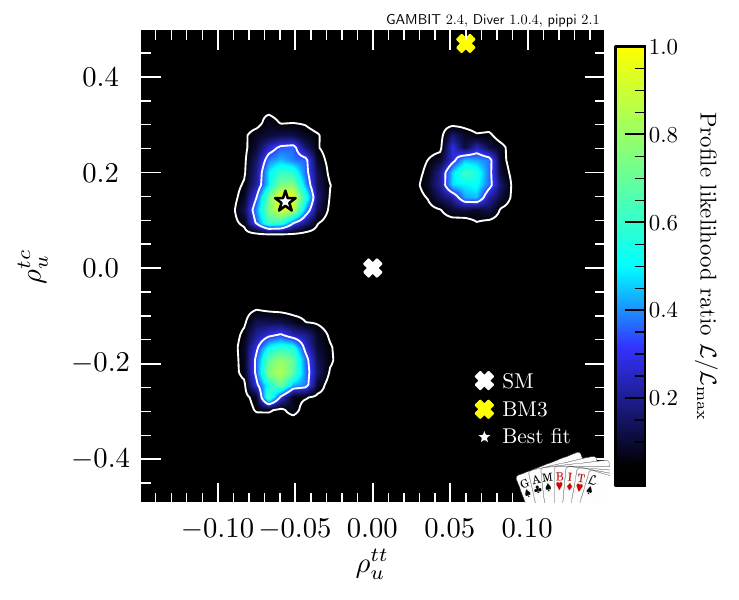}
\includegraphics[scale=0.6]{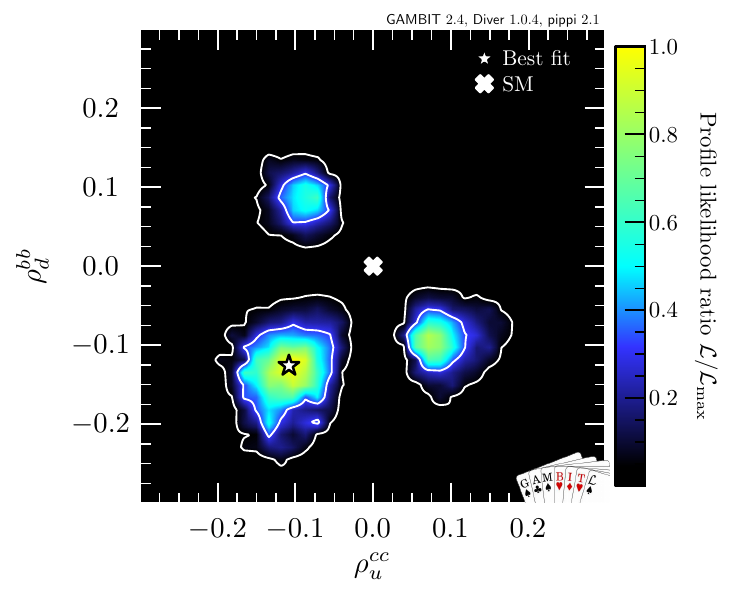}$\quad$\includegraphics[scale=0.6]
{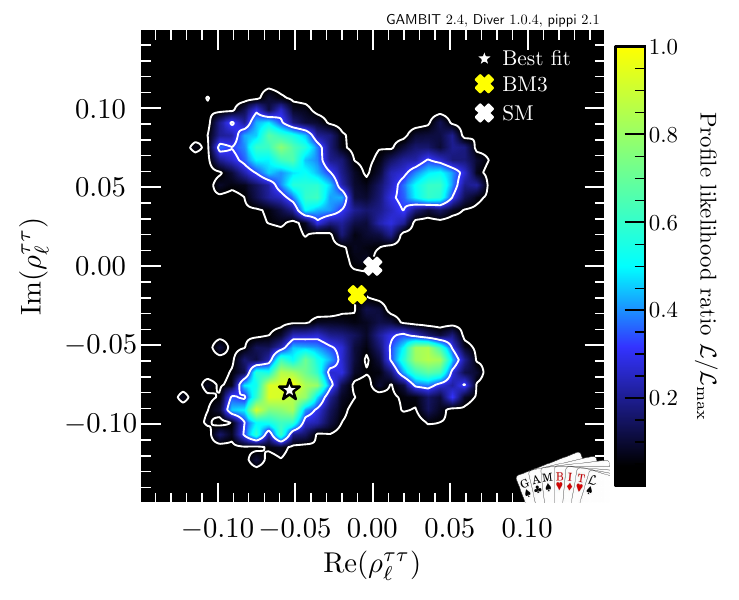}
\par\end{centering}
\caption{\emph{Profile likelihood ratio for different combinations of model parameters and observables for the full fit with the extended parameter set. As before the white star denotes the best-fit point, the white cross the SM prediction, and the yellow cross corresponds to BM3 from Ref.~\cite{Crivellin:2023sig}. White contours around the best fit point are the $1\sigma$ and $2\sigma$ confidence intervals.}\label{fig:Results_BMW}}
\end{figure}

Therefore, we now perform a more comprehensive global fit including the additional free parameters~$Y^{2,cc}_u,\,Y^{2,bb}_d$ and $Y^{2,\mu\mu}_\ell$, with the ranges described in Eq.~\eqref{eq:Ranges}. Note that the $Y^{2,ba}_f$ couplings are only used in an intermediate step as they are the pre-implemented scan parameters in \gambit. For facilitating comparison with Ref.\,\cite{Crivellin:2023sig} we show all our results in the Higgs basis, i.e. in terms of the $\rho_f^{ba}$ couplings instead, which are derived from the $Y_f^{2,ba}$ via Eq.~\eqref{eq:rhos}. In this scan we also include $b\to s\ell^+\ell^-$ data, $R_{e/\mu}$, the meson decays $\mathrm{BR}(D_{s}\to \mu\bar\nu)$ and $\mathrm{BR}(D_{s}\to \tau\bar\nu)$, the universality test $g_{\mu}/g_e$, and the new ATLAS upper limit on $t\to\mu^{+}\tau^{-}c$. For $(g-2)_\mu$ we use the SM prediction with the value of the HVP contribution obtained by BMW~\cite{Boccaletti:2024guq}. Note that this observable serves as a constraint since it agrees at the $0.9\sigma$ level with the measurement. The results of this fit are shown in Fig.~\ref{fig:Results_BMW}. The best-fit point and some predictions of selected observables can be seen in the third column of Table~\ref{tab:observables}.

From the top-right panel of Fig.~\ref{fig:Results_BMW} it can be seen that it is possible to obtain $\Delta C_9\approx-0.9$, i.e.~the best-fit point of the model-independent fit, while simultaneously explaining $R(D^{(*)})$ at the $1\,\sigma$ level (top-left) with the extended parameter set. This means that the additional Yukawa couplings, not considered in Ref.~\cite{Crivellin:2023sig}, are in fact capable of resolving the tension between $b\to s\ell^+\ell^-$ data and the rest of the observables within the global fit. The middle-left panel shows the preferred region in the $m_{A}$--$m_{H}$ plane which is sensitive to the $S$, $T$ and $U$ parameters. We can see a mild preference for a light $m_{A}$ and slightly heavier $m_{H}$ at around $200$ GeV, though degenerated masses or inverted hierarchies are also possible within $2\sigma$ of the best-fit point.\footnote{We, however, checked that the fit to all flavour and STU parameters is almost unaffected by the exchange of $m_{H}$ and $m_{A}$ which suggests there may be an undersampling of the parameter space around $m_{A}=250$ GeV.
Nevertheless we expect that this does not change our main result i.e. $\Delta C_9\approx-0.9$ is allowed in the model. }

Regarding the additional neutral scalars, ATLAS recently searched for a new particle decaying $b\bar{b}$ in top decays~\cite{ATLAS:2023mcc}. They set an upper limit on BR$(t\to c b\bar{b})\le \mathcal{O}(10^{-3})$ depending on the resonant mass. For the best-fit point we obtain BR$(t\to c A \to c b\bar{b})\approx10^{-4}$, satisfying the constraint. However, for a lighter $A$ and larger $\rho_u^{tc}$, this exotic top decay channel already probes the model. Note that the correction to oblique parameters involves all additional Higgs masses and when we set $m_{H^{\pm}}$ close to 130 GeV, this makes the values of $m_{H}$ and $m_{A}$ different compared to the case of not having the ATLAS excess. 

The disjoint preferred regions in the $\rho_{u}^{tc}-\rho_{u}^{tt}$, $\rho_{d}^{bb}-\rho_{u}^{cc}$ and $\rm{Im}\left(\rho_{\ell}^{\tau\tau}\right)-\rm{Re}\left(\rho_{\ell}^{\tau\tau}\right)$ planes (middle right and bottom panels) can be understood as the overlap of the functions associated to $\Delta C_7$, $\Delta C_9$ and $R(D^{(*)})$ defined in Eqs.~(\ref{eq:C7},\,\ref{eq:C7-1}), Eq.~(\ref{eq:C9U}) and Eq.(\ref{semileptonicWCs}), respectively. 
Even though the sign of $\rho_{u}^{tt}$ and $\rho_d^{bb}$ are not fixed by the exotic top decay, the relative sign can be fixed with flavour observables.
As a result, we get two independent regions (with opposite sign of $\rho_{u}^{tc}$) which is observed on the middle-right panel. Lastly, the bottom-right panel shows that a small but non-zero value of the imaginary part of $\rho_\ell^{\tau\tau}$ is necessary to explain $R(D^*)$ while simultaneously satisfying Higgs coupling strength constraints.

\begin{figure}[h]
\begin{centering}
\includegraphics[scale=0.6]{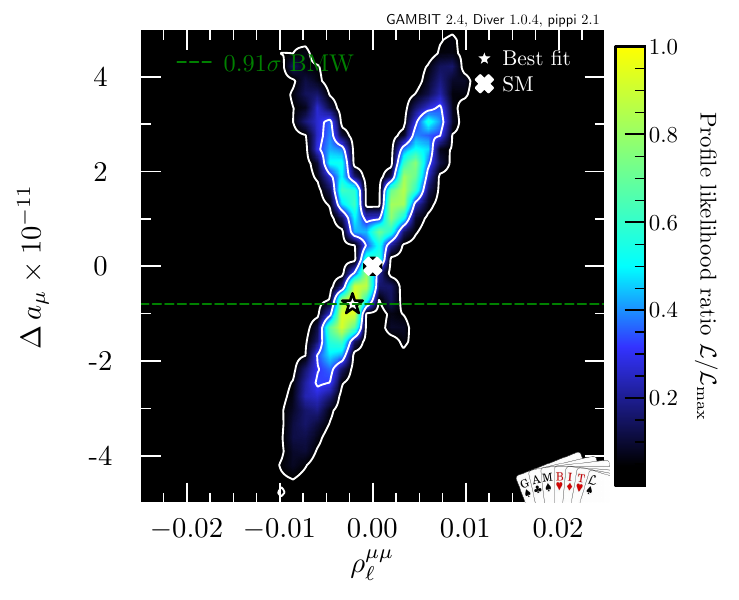}$\quad$\includegraphics[scale=0.6]
{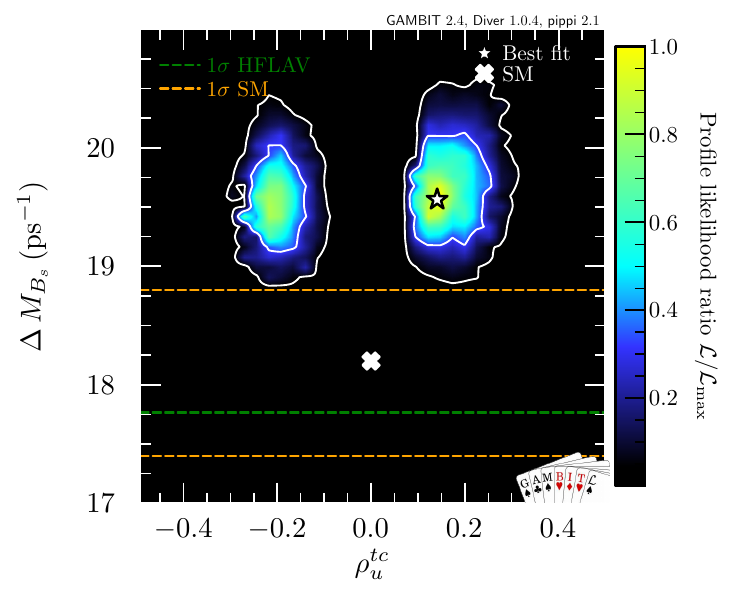}
\par\end{centering}
\caption{\emph{Profile likelihood ratios for $\Delta a_\mu$ (left) and $\Delta M_{B_s}$ (right) obtain from the full scan. As before the white star denotes the best-fit point and the white cross the SM prediction. White contours around the best fit point are the $1\sigma$ and $2\sigma$ confidence intervals.}\label{fig:params_BMW}}
\end{figure}

On the left panel of Figure~\ref{fig:params_BMW}, our fit finds a small $\rho_\ell^{\mu\mu}$ and a tiny ($\mathcal{O}(10^{-11})$) shift in the muon $g-2$. On the other hand, the right-hand panel shows that the G2HDM provides a worse fit to $\Delta M_{B_s}$ than the SM, but which is still compatible with it at the 2$\sigma$ level due to the large theoretical uncertainty of the SM (dashed orange lines). 
In the plot, the HFLAV experimental average is shown which, compared to the theory prediction, has negligible uncertainties. The fit could be improved by the inclusion of small $\rho_d^{bs}$ and $\rho_d^{sb}$ couplings which give rise to tree-level effects in $B_s-\bar B_s$ mixing.

\begin{figure}[h]
\begin{centering}
\includegraphics[scale=0.4]{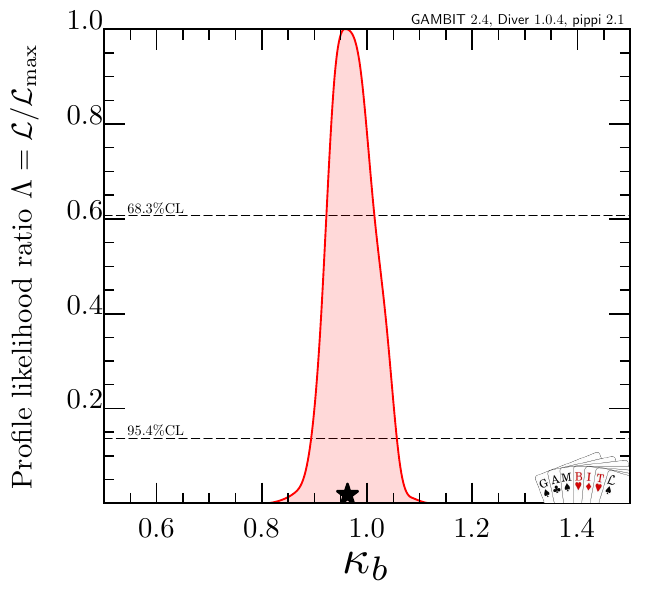}$\quad$\includegraphics[scale=0.4]
{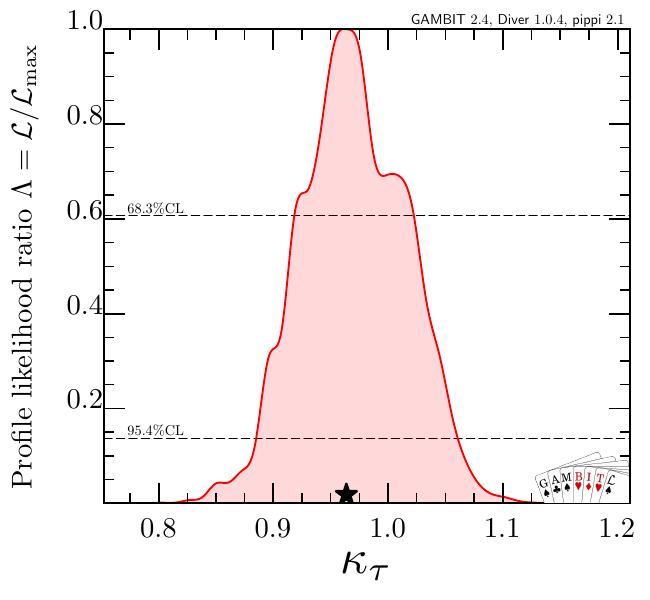}$\quad$\includegraphics[scale=0.4]
{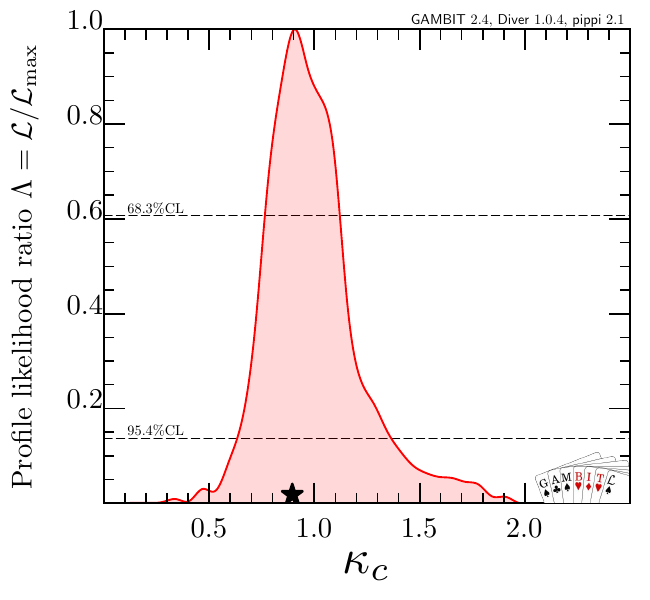}
\par\end{centering}
\caption{\emph{One-dimensional profile likelihood ratios $\mathcal{L}/\mathcal{L}_{max}$ for the Higgs coupling strength of the bottom quarks, tau lepton, and charm quarks from left to right.}\label{fig:kappas}}
\end{figure}

The predictions for the SM Higgs coupling strengths for tau leptons and charm and bottom quarks, $\kappa_{\tau,b,c}$ are shown in Figure~\ref{fig:kappas}. We see that the deviation from unity in $\kappa_\tau$ and $\kappa_b$ can be about $10\%$, and as high as $30\%$ for $\kappa_c$. Note that the current uncertainty of $\kappa_{\tau}$ and $\kappa_{b}$ is about $10\%$ but the high luminosity (HL)-LHC will shrink those uncertainties to $4\%$ and $2\%$, respectively~\cite{ATLAS:2022hsp}. On the other hand, we would need future lepton colliders to probe $\kappa_c$ at less than $10\%$~\cite{ATLAS:2022hsp,Asner:2013psa} in order to test our prediction.

\begin{figure}[h]
\begin{centering}
\includegraphics[scale=0.8]
{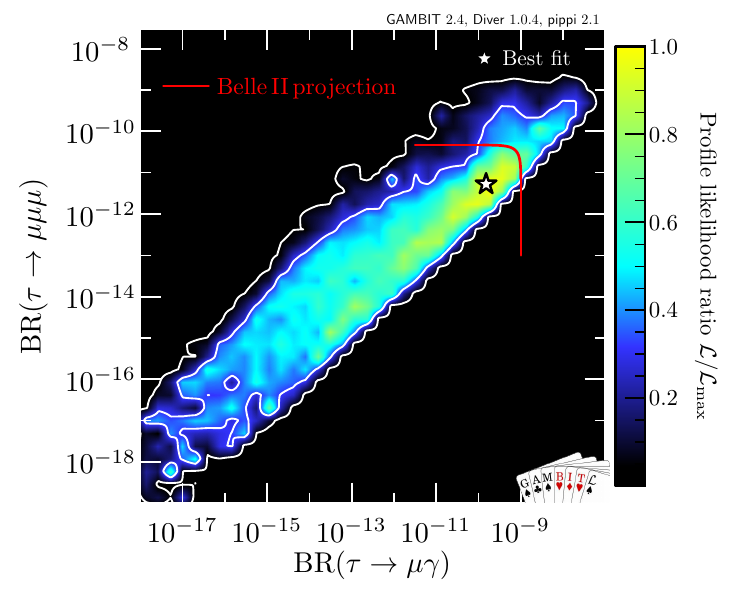}
\par\end{centering}
\caption{\emph{Profile likelihood ratios for LFV decays and projected limit from Belle-II (red solid curve). The white star denotes the best-fit point and the white contours around it are the $1\sigma$ and $2\sigma$ confidence intervals.}\label{fig:LFV_Belle2}}
\end{figure}

Figure~\ref{fig:LFV_Belle2} shows the predictions for the LFV decays $\tau\to\mu\gamma$ and $\tau\to 3\mu$ in the G2HDM. Both of the branching ratios are within $2\sigma$ of the future sensitivity limit from Belle II \cite{Belle-II:2018jsg}.\footnote{However, this is not the best possible scenario, and an even more promising projection could be obtained when using the LFV measurement obtained by the ATLAS collaboration~\cite{ATLAS:2023mvd} deviating $2.5\,\sigma$ from the SM rather than the upper limits, which we use here as a conservative approach.} Concerning LFV $B$ decays, we find that the branching ratios of $B^{(+)}\to K^{*(+)}\mu\tau$ and $B_s\to\mu\tau$ (and also for the flavour conserving $B^{(+)}\to K^{*(+)}\tau^+\tau^-$ and $B_s\to\tau^+\tau^-$) are two orders of magnitude below the future sensitivity projection from both Belle II and the HL-LHC and do not display them here.
We also note that the resulting $B_s\to \mu^{+}\mu^{-}$ branching ratio turns out to be the SM-like. Lastly, regarding the branching ratio of the $D_{s}\to \tau\bar\nu$ decay we find $\mathrm{BR}(D_{s}\to \tau\bar\nu)=5.22\times 10^{-2}$ for the best fit point value, which is within the experimental uncertainty at $1\sigma$ level.

Finally, we assess the impact of using the $W$ mass from CDF-II as input in the global fit instead of the PDG value. For this, we compare the third and fourth columns in Table \ref{tab:observables} the best-fit values using the $m_W$ from the PDG to the one using CDF-II result. We see that the NP effect in both the neutral and charged current $B$ anomalies becomes more constrained when using the CDF-II measurement compared to the PDG value. We also show in the last row of Table \ref{tab:observables} the values for the Wilks theorem ratio test $\Delta\chi^{2}=\chi^{2}_{\mathrm{SM}}-\chi^{2}_{\mathrm{G2HDM}}$, showing that a much better fit to the data is obtained for the scan using the PDG values. Notably, the CDF-II fit predicts a much smaller branching ratio for the $B_c\to\tau\bar\nu$ decay compared to others due to $|{\rm{Im}}\left(C_L^{cb}\right)|\gg |{\rm{Re}}\left(C_L^{cb}\right)|$ which needs Tera-$Z$ factories~\cite{Zheng:2021xuq,Amhis:2021cfy,Zuo:2023dzn}.

\begin{table}[h]
\centering{}\scalebox{0.98}{ %
\begin{tabular}{l|c|c|c|c}
 & BM3  & Scan 1 & PDG 2024 & CDF-II\tabularnewline
\hline 
\hline 
$m_{H^{+}}$  & $130\,{\rm {GeV}}$  & $133.7\,{\rm {GeV}}$ & $126.2\,{\rm {GeV}}$  & $133\,{\rm {GeV}}$ \tabularnewline
$m_{H,A}$  & $200\,{\rm {GeV}}$  & $181.7,\,184.2\,{\rm {GeV}}$ & $205,\,158\,{\rm {GeV}}$  & $227,\,201\,{\rm {GeV}}$\tabularnewline
$c_{\beta\alpha}$  & $0.1$  & $0.019$ & $0.007$ & $0.03$\tabularnewline
$\rho_{u}^{tt}$    & $0.06$  & $0.06$ & $-0.06$ & $-0.06$\tabularnewline
$\rho_{u}^{tc}$   & $0.47$  & $0.23$  & $0.14$ & $-0.15$\tabularnewline
$\rho_{u}^{cc}$ & - & -  & $-0.1$ & $0.1$\tabularnewline
$\rho_{d}^{bb}$  & - & - & $-0.07$ & $0.1$\tabularnewline
$\rho_{\ell}^{\tau\tau}$  & $-0.01(1\pm 1.8i)$  & $-0.03(1\pm1.5\,i)$  & $-0.05(1\pm1.6\,i)$  & $-0.002(1\pm25\,i)$\tabularnewline
$\rho_{\ell}^{\mu\mu}$ & - & - & $-0.0015$ & $0.002$\tabularnewline
$\rho_{\ell}^{\mu\tau}$  & $0.01$  & $5\times 10^{-4}$ & $0.003$ & $-0.001$\tabularnewline
$\rho_{\ell}^{e\tau}$   & $0.006$  & $0.008$ & $3\times 10^{-4}$ & $0.007$\tabularnewline
\hline 
${\rm {BR}}(t\to b\bar{b}c)$  & $0.163\%$  & $0.157\%$ & $0.156\%$  & $0.157\%$\tabularnewline
${\rm {BR}}(h\to\mu\tau)$  & $0.077\%$  & $6.7\times 10^{-8}$ & $3.5\times 10^{-7}$ & $6.8\times 10^{-7}$\tabularnewline
${\rm {BR}}(h\to e\tau)$   & $0.028\%$,  & $2.1\times10^{-5}$ & $3.6\times 10^{-9}$ & $2.8\times 10^{-5}$\tabularnewline
$R(D)$   & $0.357$  & $0.350$ & $0.346$ & $0.371$\tabularnewline
$R(D^{*})$  & $0.271$  & $0.276$ & $0.277$ & $0.258$\tabularnewline
${\rm {BR}}(\mu\to e\gamma)$  & $2.2\times10^{-13}$  & $1.7\times10^{-15}$ & $7.6\times10^{-17}$ & $3.1\times10^{-15}$\tabularnewline
$R_{B_{s}}$ & $0.002$  & $-0.005$ & $0.075$ & $0.062$ \tabularnewline
${\rm {BR}}(B_{c}\to\tau\bar{\nu})$   & $30\,\%$  & $39\,\%$ & $40\,\%$  & $8\,\%$\tabularnewline
${\rm {BR}}(t\to ch)$  & $3.1\times10^{-4}$  & $2.4\times10^{-6}$ & $1.4\times10^{-7}$ & $2.2\times10^{-6}$\tabularnewline
$\Delta C_{9}$  & $-0.47$  & $-0.072$ & $-0.83$ & $-0.76$\tabularnewline
$\Delta C_{7}$  & $-0.035$  & $-0.015$ & $-0.016$ & $-0.011$\tabularnewline
$\kappa_{\tau}$  & $0.91$ &  $0.95$ & $0.97$ & $1.00$\tabularnewline
$\Delta a_{\mu}^{BMW}$ & - & - & $-0.8\times10^{-11}$ & $1.2\times10^{-11}$\tabularnewline
$STU$ ($\Delta\chi^{2})$   & $-2.5$  & $0.014$ & $-0.06$ & $-11.5$ \tabularnewline
\hline
Total $\Delta \chi^2$ & $2.20$  & $23.83$ & $81.07$ & $64.30$ \tabularnewline
\end{tabular}} \caption{\emph{The value of the parameters for BM3 (first column), the best-fit point of the first scan (second column), the best-fit point of the second scan with using the PDG value (third column) or the CDF-II value (fourth column) for the value of $m_W$. The corresponding predictions for various observables are shown, where relevant. Here we define $\Delta\chi^{2}=\chi^{2}_{\mathrm{SM}}-\chi^{2}_{\mathrm{G2HDM}}$.}
}
\label{tab:observables} 
\end{table}

\section{Conclusions}\label{sec:Conclusions}

Explaining the anomalies in semi-leptonic $B$ decays remains a challenge for model building. In fact, after the disappearance of deviations from unity in the ratios $R(K^{(*)})$ testing lepton flavour phenomenology, this has become even more difficult and fewer NP models remain valid~\cite{Capdevila:2023yhq}. One of them is the 2HDM with a generic flavour structure. We show that it is possible to describe both the charged and neutral current anomalies in semi-leptonic $B$ decays at the $1\,\sigma$ level within the G2HDM, while satisfying the experimental constraints.  We do this by performing a global fit via \gambit that includes the constraints from all other flavour observables, top decays and electroweak precision observables. For the latter, we used the $W$ mass from the PDG (which does not include the CDF-II measurement) and the SM prediction for $(g-2)_\mu$ with the HVP contribution calculated by the BMW collaboration, which both imply good agreement between SM and experiment so that these observables act as constraints on new physics.  

We stress that the value used for the HVP contributions to $(g-2)_\mu$ plays a crucial role here. Specifically, using the BMW calculation, which implies a SM prediction close to the measured $(g-2)_\mu$ value, allows a simultaneous explanation of $R(D^{(*)})$ and $b\to s\ell^+\ell^-$ data.  If instead the SM $(g-2)_\mu$ value of the White Paper is used without updating the HVP contributions, the combined fit can no longer describe both $B$ anomalies while generating a larger $\Delta a_{\mu}$ of about $1\times10^{-9}$ (which is still significantly smaller than the White Paper prediction). Similarly, if the PDG value for the $W$ mass is replaced by the CDF-II measurement, the model could not fit the oblique parameters while at the same time improving the fit to the semi-leptonic $B$ anomalies as much as before.  

Interestingly, we found that if we do not include $b\to s\ell^+\ell^-$ in the fit but predict it from the other observables within the G2HDM, a negative value of $\Delta C_9$, as suggested by global model-independent fits, is predicted, even though the absolute value is smaller than what is preferred by data. In all scenarios describing the $B$ anomalies at the $1\sigma$ level, we find that $\Delta M_{B_s}$ is in worse agreement with data than the SM, even though still compatible due to the large theoretical uncertainty. Finally, our model can be tested at Belle II in LFV searches for $\tau\to3\mu$ and $\tau\to\mu\gamma$ and measurements of $\kappa_{b}$ and $\kappa_{\tau}$ in the future HL-LHC.

\acknowledgments

We thank Douglas Jacob for discussions in the early stages of
the project. The work of both PA and CS is supported by the National Natural Science Foundation of China (NNSFC) under grant No.~12150610460.  AC is supported by a professorship grant from the Swiss National Science Foundation (No.\ PP00P2\_211002). The work of CS is also supported by the Excellent Postdoctoral Program of Jiangsu Province grant No.~2023ZB891 and the work of PA by NNSFC Key Projects grant No.~12335005 and the supporting fund for foreign experts grant wgxz2022021. TEG acknowledges funding by the Deutsche Forschungsgemeinschaft (DFG) through the Emmy Noether Grant No.~KA 4662/1-2. SI enjoys support from JSPS KAKENHI Grant Number 24K22879 and JPJSCCA20200002. We acknowledge the EuroHPC Joint Undertaking for awarding this project access to the EuroHPC supercomputer LUMI, hosted by CSC (Finland) and the LUMI consortium through a EuroHPC Extreme Scale Access call.

\appendix

\section{Barr-Zee diagrams for $(g-2)_\mu$ in the G2HDM}\label{sec:BZ-G2HDM}

\begin{figure}[h]
	\centering
    \includegraphics[scale=0.8]{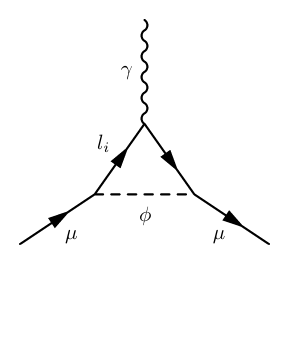}
    \includegraphics[scale=0.8]{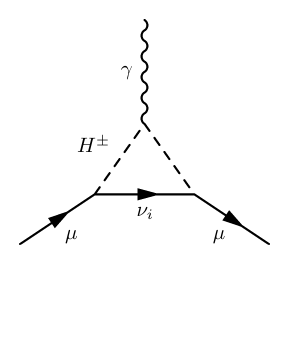}
	\caption{\emph{One-loop diagrams that contribute to $(g-2)_\mu$ involving a neutral (charged) scalar diagram on the left (right).  The index $i=e,\mu,\tau$ and $\phi=h,H,A$ are defined.  \label{fig:gm2OneLoop}}}
\end{figure}

\begin{figure}[h]
	\centering
    \includegraphics[scale=0.8]{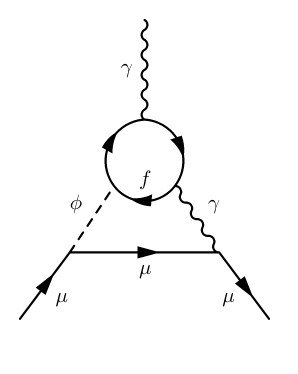}
    \includegraphics[scale=0.8]{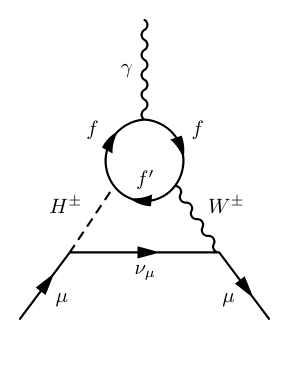}
    \includegraphics[scale=0.8]{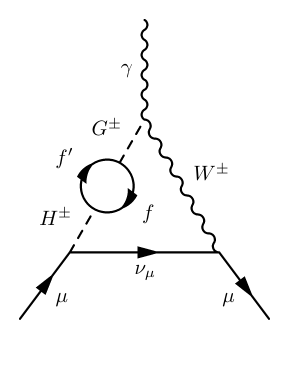}
	\caption{\emph{Two-loop fermionic Barr-Zee diagrams that contribute to $(g-2)_\mu$. The internal photon $\gamma$ may be replaced by a $Z$ boson, $f'$ is the $SU(2)_L$ partner of $f$ and $\phi=h,H,A$.  \label{fig:gm2TwoLoopFermionic}}}
\end{figure}

\begin{figure}[h]
	\centering
    \includegraphics[scale=0.8]{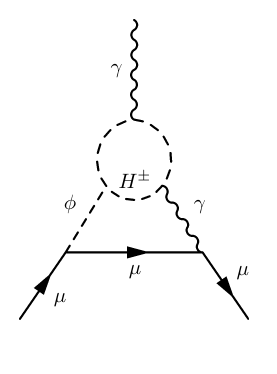}
    \includegraphics[scale=0.8]{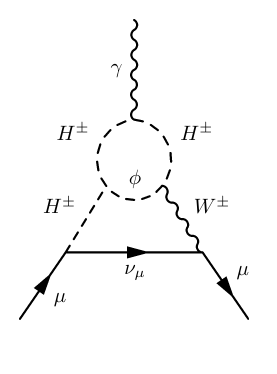}
    \includegraphics[scale=0.8]{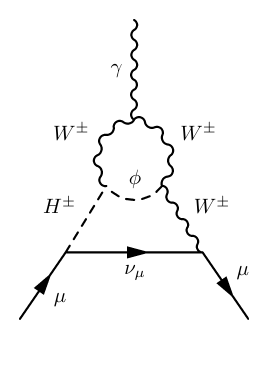}
	\caption{\emph{Two-loop bosonic Barr-Zee diagrams contributing to $(g-2)_\mu$.  Here $\phi=h,H,A$, as in Figure \ref{fig:gm2OneLoop}.  In the left panel, the internal photon $\gamma$ may be replaced by a $Z$ boson, and the internal $H^\pm$ with a $W^\pm$ boson.  \label{fig:gm2TwoLoopBosonic}}}
\end{figure}

\begin{figure}[h]
	\centering
    \includegraphics[scale=0.8]{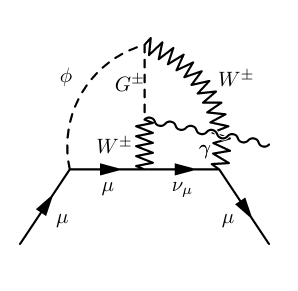}
    \includegraphics[scale=0.8]{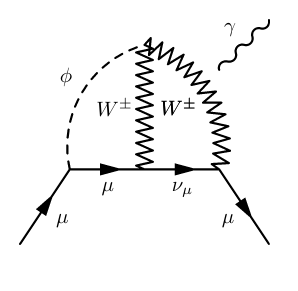}
    \includegraphics[scale=0.8]{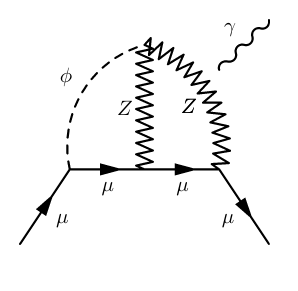}
	\caption{\emph{Two-loop bosonic 3 boson diagrams contributing to muon $g-2$.  Here $\phi=h,H,A$ as in Figure \ref{fig:gm2OneLoop}.  In the middle and right panels, the detached photon could be attached to either of the vector boson lines in the diagram.  In the right panel, the scalar boson $\phi$ may swap positions with the $Z$ bosons.  \label{fig:gm2TwoLoop3Boson}}}
\end{figure}

We divide the leading contributions to $\Delta\,a_{\mu}$ in the G2HDM into three groups
\begin{equation}
\Delta\,a_{\mu}^\textrm{BSM} = \Delta\,a_{\mu}^{\mathrm{1L}} + \Delta\,a_{\mu}^F + \Delta\,a_{\mu}^B,
\end{equation}
which correspond to the one-loop, two-loop fermionic, and two-loop bosonic contributions, respectively.  These are shown in Figures~\ref{fig:gm2OneLoop}-\ref{fig:gm2TwoLoop3Boson} and the corresponding expressions for the G2HDM are given in Refs.~\cite{Athron:2021evk,Athron:2021auq}. 

The two-loop fermionic Barr-Zee diagrams are shown in Figure~\ref{fig:gm2TwoLoopFermionic} and can be divided into neutral and charged contributions
\begin{equation} \label{eqn:twoloopfermionic}
    \Delta\,a_{\mu}^F = \Delta\,a_{\mu}^{F,{\rm neutral}} + \Delta\,a_{\mu}^{F,{\rm charged}}.
\end{equation}

The two-loop bosonic contributions can be split up further into  three groups
\begin{equation} \label{eqn:twoloopbosonic2}
   \Delta\,a_{\mu}^B = \Delta\,a_{\mu}^\textrm{B, EW add} + \Delta\,a_{\mu}^\textrm{B, Yuk} + \Delta\,a_{\mu}^\textrm{B, non-Yuk}.  
\end{equation}
The first term $\Delta\,a_{\mu}^\textrm{B,EW add}$ represents BSM contributions from two-loop bosonic diagrams where only SM particles and the SM-like Higgs boson $h$ appear in the loops, i.e.\ the left panel of Figure \ref{fig:gm2TwoLoopBosonic} with $\phi=h$ and $H^\pm$ replaced with $W^\pm$ and the diagrams in Figure \ref{fig:gm2TwoLoop3Boson} with $\phi=h$.  BSM effects enter here because in the G2HDM $h$ can have non-SM effects. However to avoid double counting and get only the BSM contribution we must subtract from this the SM value of these diagrams.  The second term $\Delta\,a_{\mu}^\textrm{B, Yuk}$ includes all diagrams that involve BSM fields with a Yukawa coupling, e.g. the $\phi\ne h$ versions of both the diagram in the left panel of Figure \ref{fig:gm2TwoLoopBosonic} and the diagrams in Figure \ref{fig:gm2TwoLoop3Boson}.  The last term $\Delta\,a_{\mu}^\textrm{B, non-Yuk}$ represents two-loop bosonic diagrams that do not involve Yukawa couplings, such as those in the middle and right panels in Figure \ref{fig:gm2TwoLoopBosonic}.  
The definitions of the contributions $\Delta\,a_{\mu}^\textrm{B, EW add}$ and $\Delta\,a_{\mu}^\textrm{B, non-Yuk}$ are shown in Eqs.\ (49) and (71) respectively in Ref.\ \cite{Cherchiglia:2016eui}, while the Yukawa contribution $\Delta\,a_{\mu}^\textrm{B, Yuk}$ is shown in Eq.\ (67) of Ref.\ \cite{Athron:2021evk}.

\section{Fit with $(g-2)_\mu$ value from White Paper}
\label{sec:wp}

\begin{figure}[h]
\begin{centering}
\includegraphics[scale=0.4]{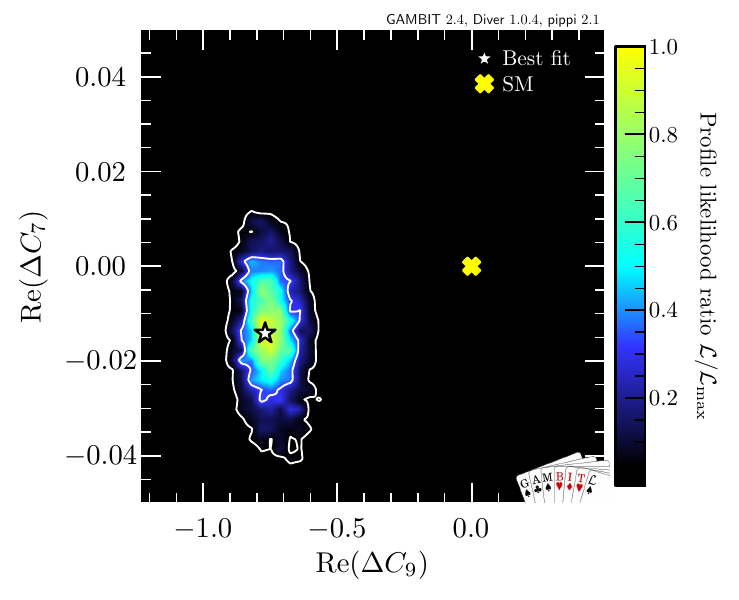}$\quad$\includegraphics[scale=0.4]{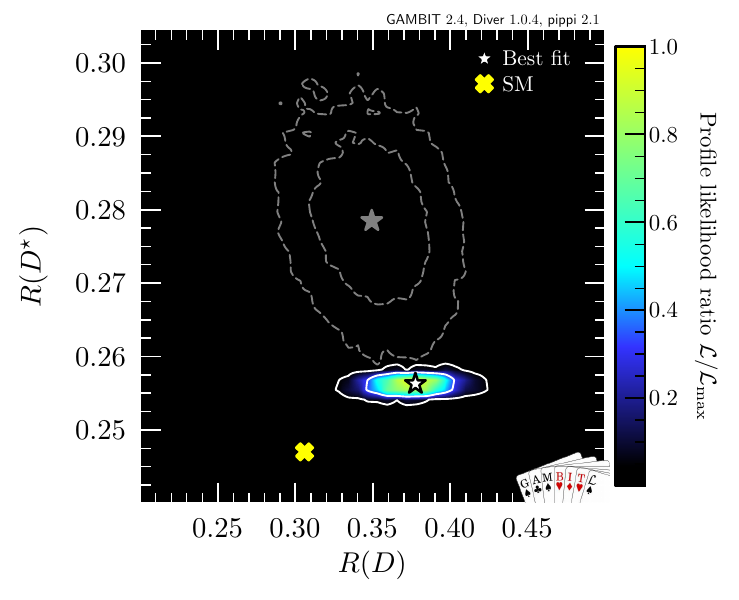}$\quad$\includegraphics[scale=0.4]
{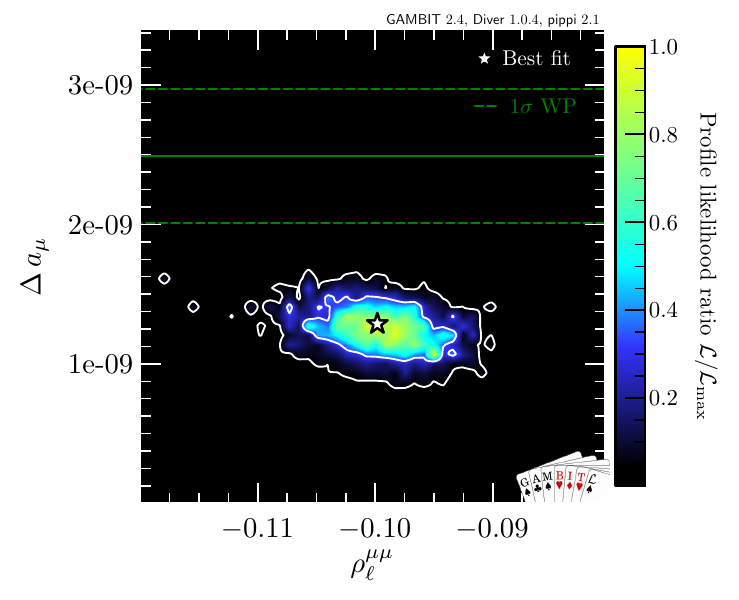}

\par\end{centering}
\caption{\emph{Profile likelihood ratios $\mathcal{L}/\mathcal{L}_{max}$ for different
2D plots of the parameter space for the scan using the WP value for muon $g-2$. The grey contours on the middle panel are the comparison with respect to the previous scan in section \ref{sec:Results}.}\label{fig:Results_WP}}
\end{figure}

Here we present the results of a scan using the WP value~\cite{Aoyama:2020ynm} for the SM prediction for the muon $g-2$ instead of the one where BMW is used for HVP. We find $|\rho_\ell^{\tau\tau}|\ll|\rho_u^{cc}|$ resulting in BR($\phi\to \tau\bar{\tau}$) $\ll$ BR($\phi\to c\bar{c}$) and hence multi lepton search and chargino-neutralino searches would be less relevant.
In Figure~\ref{fig:Results_WP} we can see that it is possible to simultaneously fit all observables only at the $2\,\sigma$ level, with the exception of $R(D^*)$ which is SM-like within this global fit. This is expected due to the smaller experimental uncertainty for the WP value (compared to the BMW one), strongly constraining the $\rho^{tc}_{u}$ Yukawa coupling, although requiring a large $\rho_u^{cc}\approx-0.5$, in possible conflict with $pp\to c\bar{c}\to \phi$ and $pp\to cs \to H^\pm$ searches. Finally, we find that if using the CDF-II value for $m_W$ the model is ruled out at the $2\sigma$ level for an explanation of the WP value in the G2HDM.

\bibliographystyle{JHEP}
\bibliography{GTHDM}

\end{document}